\documentclass[12pt]{report}
\usepackage[polutonikogreek, english]{babel}
\usepackage[latin1]{inputenc}
\usepackage{amsthm,amsfonts}
\usepackage{amsmath}
\usepackage{setspace}
\input xy
\xyoption{all}

\begin{document}

\pagenumbering {roman}

\begin{titlepage}
\begin{center}
{\large \sffamily{{Doctoral Thesis}}}\\
\vspace{3cm}
{\Huge \sffamily Hopf Algebras in\\ Deformed Quantum Theories}

\vspace{5cm}
{\Large \sffamily {Paulo Guilherme Castro} }\\
\vspace{7cm}
{\large  \sffamily  Centro Brasileiro de Pesquisas F\'isicas (CBPF)} \\
{\large  \sffamily  Rio de Janeiro, August 2010 }

\vfill

\vspace{0.5cm}

\vskip -0.5cm

\end{center}
\end{titlepage}

\newpage
\bigskip
\onehalfspacing
\begin{flushright}
{\selectlanguage{polutonikogreek} \large{To~is zhto~us`i ti pr~agma \char162\hspace{1pt}  eVuresin \char226pakolouje~in e\char202k`os  \char162\hspace{1pt}  \char138rnhsin e\char205r'esews ka`i  \char130katalhy'ias \char229molog'ian \char162\hspace{1pt} \char226pimon`hn zht'hsews.}}
\vspace{1.2cm}

Sextus Empiricus
\end{flushright}
\vspace{8cm}

\begin{center}
To Isabel
\end{center}

\vspace{1.2cm}

\addtocounter{page}{1}
\singlespacing
\tableofcontents

\newpage
\addcontentsline{toc}{section}{Acknowledgments}
\chapter*{Acknowledgments}
\onehalfspacing

\hspace{0.5cm} To my advisor Prof. Francesco Toppan, for his availability and generosity, and for the many things he taught me.

To Profs.\ Biswajit Chakraborty and Zhanna Kuznetsova, for the profitable scientific collaboration. 

To my colleagues, particularly Ricardo Kullock, Rodrigo Maier and Thiago Guerreiro, for their friendship and the enjoyable exchange of ideas. 

To my friends and family, for their unconditional support.

To the CBPF staff, for their helpfulness.

The author received financial support from CNPq.

\newpage


\addcontentsline{toc}{section}{Abstract}
\chapter*{Abstract}

In this work we apply the Drinfel'd twist of Hopf algebras to the study of deformed quantum theories.

We prove that, by carefully considering the role of the central extension, it is indeed possible to construct the universal enveloping algebra of the Heisenberg algebra and deform it by means of a Drinfel'd twist, which yields a noncommutative theory. Furthermore, we show that in the second-quantization formalism the Hopf algebra structure of the Heisenberg algebra (both undeformed and deformed) can be obtained from the Hopf algebra of the Schr\"odinger fields and oscillators, as long as they are taken to be odd generators of the $osp(1|2)$ superalgebra.

We study the deformation of the fermionic Heisenberg algebra and present an identification with the algebra of the one-dimensional ${\cal N}$-extended supersymmetric quantum mechanics, possible for even ${\cal N}$. A second construction for the deformation of the one-dimensional ${\cal N}$-extended supersymmetric quantum mechanics is presented in the superspace representation, where the supersymmetry generators are realized in terms of operators belonging to the universal enveloping superalgebra of one bosonic and several fermionic oscillators. In both constructions we recover, in a more general setting, some Cliffordization results of the literature.

\newpage

\pagenumbering {arabic}

\addcontentsline{toc}{chapter}{Introduction}
\chapter*{Introduction}
\doublespacing

The problem of the divergences has afflicted quantum field theory since its early times \cite{born}. The possibility of solving this problem by introducing a fundamental length as a natural ultraviolet regulator was first proposed by Heisenberg \cite{heisenberg}. The idea (inspired in the position-momentum uncertainty relation) was that below the elementary distance the concepts of point and instant would no longer make sense, and would be replaced by the notion of a fuzzy space-time. The simplest way of implementing space-time noncommutativity is by means of the relation 
$$
[x_\mu,x_\nu]=i\theta_{\mu\nu},
$$
where $\theta_{\mu\nu}$ is a constant skew-symmetric matrix. These commutation relations induce the uncertainty relation 
$$
\Delta x_\mu\Delta x_\nu\geq\frac{1}{2}|\theta_{\mu\nu}|,
$$
with $|\theta_{\mu\nu}|$ of the order of magnitude of the square of the fundamental length. 

A natural candidate for this length (see \cite{dfr}, \cite{dfr2}) is the so-called \emph{Planck length}
$$
\ell_P=\sqrt{\frac{G\hbar}{c^3}},
$$
which combines the fundamental constants of nature (Newton's constant $G$, Planck's constant $\hbar$ and the speed of light $c$) in a dimensionally appropriate way. Because it is a constant of nature, and not an artificially imposed cutoff, such an ultraviolet regulator would be extremely welcome in quantum theory. 

However, one can immediately see that a theory of this kind is not Lorentz covariant. Indeed, the coordinates  $x_\mu$ transform as vectors, while $\theta_{\mu\nu}$ is constant in all reference frames. 

To avoid this inconvenience, Snyder introduced in \cite{snyder} the commutation relations
$$
[x_\mu,x_\nu]=i\theta(x_\mu p_\nu-x_\nu p_\mu),
$$
which are clearly covariant under Lorentz transformations. However, probably due to the failure at making accurate experimental predictions and to the great success of the renormalization techniques, this proposal received little attention at the time, and quantum theory in noncommutative space-time went through a long period of ostracism.

The interest in this kind of theory resurfaced with Seiberg and Witten \cite{sw}, who showed that string theory, at a certain low-energy limit, can be realized as an effective quantum field theory in a noncommutative space-time.  Some important mathematical developments of the 1980s have also contributed to this rebirth, and noncommutative theories have been an area of intense research since then. For a review of the topic, see e.g. \cite{dn}, \cite{szabo} and \cite{bisw}.

Among the mathematical developments mentioned above, the most important was probably the introduction of the notion of quantum group by Drinfel'd (\cite{drin}, \cite{drin85}) and Jimbo \cite{jimbo}, initially in the context of quantum integrable systems. The idea is that, if the structure of space-time is deformed, the symmetry groups that act on it must also be deformed. However, Lie groups and Lie algebras, which are the objects that implement symmetries, are said to be rigid, i.e., not susceptible to deformation.  Therefore, to put into practice the program of deforming symmetry groups, it was necessary to resort to Hopf algebras, introduced much earlier within the context of algebraic topology \cite{hopf} and object of interest of mathematicians since then (see the classical \cite{sweedler} and \cite{abe} and the more recent \cite{das} and \cite{balhopf}). 

Hopf algebras provide the suitable mathematical framework for the study of quantum groups. The topic is immensely rich and   
is well exposed, for instance, in \cite{chari} and \cite{majid}. For some interesting pioneering works, see \cite{jimbo2}. 

This formalism allowed, for example, to reconcile noncommutativity and Lorentz covariance: in \cite{cknt}, the problem is solved by considering the underlying commutative (therefore manifestly Lorentz covariant) space-time endowed with a deformed product which is implemented by means of a Drinfel'd twist. 

In this work, we will apply the machinery of Hopf algebras and Drinfel'd twist to the bosonic and fermionic Heisenberg algebras. In chapter \textbf{1}, we make a succint exposition of the basic general theory of Hopf algebras and Drinfel'd twist, and particularly of how to apply it to the universal enveloping algebra of a Lie algebra. In chapter \textbf{2}, we initially show that it is indeed possible to construct the universal enveloping algebra of the Heisenberg algebra, and then proceed to study its deformation both in the context of first quantization and in the context of second quantization, where it is realized by integrated bilinears of the Schr\"odinger fields and oscillators. This requires the use of Wigner oscillators. In chapter \textbf{3}, we study the deformation of the fermionic Heisenberg algebra and present an identification with the algebra of the one-dimensional  $\mathcal{N}$-extended supersymmetric quantum mechanics, possible for even values of $\mathcal{N}$. We also study the deformation of the one-dimensional  $\mathcal{N}$-extended supersymmetric quantum mechanics in its superspace representation (possible for all $\mathcal{N}$), recovering, in a more general setting, some results of the previous literature on nonanticommutative theories with violation of the Leibniz rule (\cite{as}, \cite{krt}).

The original results of this thesis can be found in \cite{cct} and \cite{cckt}.

\chapter{Hopf Algebras and Drinfel'd Twist}

In this chapter we will briefly review the mathematical structures which will be necessary to the understanding of the following chapters. 

\section{Hopf Algebras}
Let $A$ be a vector space over the field $\mathbf{k}$ and $\mu:A\otimes A\rightarrow A$ and $\eta:\mathbf{k}\rightarrow A$ $\mathbf{k}$-linear maps. We call $(A,\mu,\eta)$, or simply $A$, a \emph{unital associative algebra} if the diagrams
\[
\centerline{
\xymatrix{
A\otimes A\otimes A
\ar[r]^{id\otimes\mu}
\ar[d]_{\mu\otimes id}
& A\otimes A
\ar[d]_{\mu}\\
A\otimes A
\ar[r]^{\mu}
& A
}
}
\]
and
\[
\centerline{
\xymatrix{
\mathbf{k}\otimes A
\ar[r]^{\eta\otimes id}
\ar[d]
& A\otimes A
\ar[dl]_{\mu}\\
A
& A\otimes\mathbf{k}
\ar[l]
\ar[u]_{id\otimes\eta}
}
}\]
commute.

The map $\mu$ is called \emph{multiplication} and the map $\eta$ \emph{unit}, and $\mu$ and $\eta$ are called \emph{structures} or \emph{structural maps} of the algebra. We denote $\mu(a\otimes b)=a\cdot b$ ($a,b\in A$). The first diagram corresponds to the property of associativity, which can also be expressed as
\begin{equation}
    (a\cdot b)\cdot c=a\cdot(b\cdot c),
\end{equation}
$a,b,c\in A$.

The second diagram guarantees the existence of the left and right unit  $\mathbf{1}$ in $A$ , with $\eta(1)=\mathbf{1}$  (1 is evidently the unit in $\mathbf{k}$).

The notion of \emph{coalgebra} can be naturally introduced dualizing (in a category theoretical sense \cite{borceux}) the  definitions above, i.e., the direction of arrows must be inverted. Consider a vector space $C$ over the field $\mathbf{k}$ and the $\mathbf{k}$-linear maps $\Delta:C\rightarrow C\otimes C$ and $\epsilon:C\rightarrow\mathbf{k}$. We call $(C,\Delta,\epsilon)$, or for simplicity $C$, a \emph{coalgebra} if the diagrams
\[
\centerline{
\xymatrix{
C\otimes C\otimes C
& C\otimes C
\ar[l]_{id\otimes\Delta}\\
C\otimes C
\ar[u]_{\Delta\otimes id}
& C
\ar[u]^{\Delta}
\ar[l]^{\Delta}
}
}\]
and
\[
\centerline{
\xymatrix{
\mathbf{k}\otimes C
& C\otimes C
\ar[l]_{\epsilon\otimes id}
\ar[d]^{id\otimes\epsilon}
\\
C
\ar[u]
\ar[r]
\ar[ru]^{\Delta}
& C\otimes\mathbf{k}
}
}\]
commute.

The map $\Delta$ is called \emph{coproduct} or \emph{comultiplication} and the map $\epsilon$ \emph{counit}, and  they are known as \emph{costructures} or \emph{costructural maps} of the coalgebra. It is convenient here to introduce the \emph{Sweedler notation} \cite{sweedler}, which consists in suppressing the summation indices in the expression of the coproduct, i.e.,
\begin{equation}
    \Delta(a)=\sum_i \left(a_1\right)_i\otimes \left(a_2\right)_i \equiv a_1\otimes a_2,
\end{equation}
$a\in C$.

The property encoded in the first diagram is called coassociativity and is equivalent to the expression
\begin{equation}
    \Delta(a_1)\otimes a_2=a_1\otimes\Delta(a_2),
\end{equation}
while the property of counitarity contained in the second diagram is guaranteed by $\epsilon(\mathbf{1})=1$, where $\mathbf{1}$ is the unit in $C$ and $1$ the unit in $\mathbf{k}$, and the expression
\begin{equation}
\epsilon(a_1)\otimes a_2 = a_1\otimes\epsilon(a_2)
\end{equation}
holds $\forall a\in C$.

We employ the notation $\mu_A$, $\eta_A$, $\mathbf{1}_A$ and $\Delta_C$, $\epsilon_C$ when several algebras and coalgebras are involved, omitting the subscript when there is no risk of confusion.

Let now $A$ and $\tilde{A}$ be algebras and $h:A\rightarrow \tilde{A}$ a linear map. We say that $h$ is an \emph{algebra homomorphism} if it is multiplicative and preserves unit, i.e., the diagrams 
\[
\centerline{
\xymatrix{
A\otimes A
\ar[r]^{h\otimes h}
\ar[d]_{\mu_A}
& \tilde{A}\otimes \tilde{A}
\ar[d]^{\mu_{\tilde{A}}}
 \\
A
\ar[r]^{h}
& \tilde{A}
}
}\]
and
\[
\centerline{
\xymatrix{
A
\ar[rr]^{h}
&& \tilde{A}
 \\
& \mathbf{k}
\ar[lu]^{\eta_A}
\ar[ru]_{\eta_{\tilde{A}}}
 }
 }\] commute.

These conditions can be expressed respectively as
\begin{eqnarray}
  h\left(a\cdot b\right) &=& h(a) \tilde{\cdot} h(b) \\
  h\left(\mathbf{1}_A\right) &=& \mathbf{1}_{\tilde{A}},
\end{eqnarray}
where $a,b\in A$, $\cdot$ is the multiplication in $A$ and $\tilde{\cdot}$ the multiplication in $\tilde{A}$.

Analogously, for $C$ and $\tilde{C}$ coalgebras, a linear map $g:C\rightarrow \tilde{C}$ is said to be a \emph{coalgebra homomorphism} if it is comultiplicative  and preserves counit, i.e., the diagrams
\[\centerline{
\xymatrix{
C\otimes C
\ar[r]^{g\otimes g}
& \tilde{C}\otimes \tilde{C}
 \\
C
\ar[r]^{g}
\ar[u]^{\Delta_C}
& \tilde{C}
\ar[u]_{\Delta_{\tilde{C}}}
}
}\]
and
\[\centerline{
\xymatrix{
C
\ar[rr]^{g}
\ar[rd]_{\epsilon_C}
&& \tilde{C}
\ar[ld]^{\epsilon_{\tilde{C}}} \\ 
& {\bf k}
}
}\]
commute.

The information encoded in the diagrams can be written simply as
\begin{eqnarray}
  \Delta_{\tilde{C}}(g(c)) &=& g(c_1)\otimes g (c_2) \\
  \epsilon_{\tilde{C}} (g(\mathbf{1}_C)) &=& 1,
\end{eqnarray}
where $c\in C$ and $c_1\otimes c_2=\Delta_C(c)$.

Let now $(B,\mu, \eta)$ be an algebra over $\mathbf{k}$ and $(B, \Delta, \epsilon)$ a coalgebra over $\mathbf{k}$. We call $(B,\mu,\eta,\Delta,\epsilon)$, or simply $B$, a \emph{bialgebra} if the structures $\mu$ and $\eta$ and the costructures $\Delta$ and $\epsilon$ are \emph{compatible}, that is, $\mu$ and $\eta$ are coalgebra homomorphisms and $\Delta$ and $\epsilon$ are algebra homomorphisms.

Compatibility among structures and costructures can be diagramatically expressed as the commutativity of the diagrams 
\[
\centerline{
\xymatrix{
B\otimes B \ar[r]^{\mu} \ar[d]_{\Delta\otimes \Delta}
& B \ar[r]^{\Delta }
& B\otimes B  \\
B\otimes B\otimes B\otimes B \ar[rr]_{id\otimes\tau\otimes id}
&& B\otimes B\otimes B\otimes B,\ar[u]_{\mu\otimes\mu}
}
}\]

\[
\centerline{
\xymatrix{
B
\ar[r]^{\Delta}
& B\otimes B
 \\
\mathbf{k}
\ar[r]
\ar[u]^{\eta}
& \mathbf{k}\otimes\mathbf{k},
\ar[u]_{\eta\otimes\eta}
}
}\]

\[
\centerline{
\xymatrix{
B\otimes B
\ar[r]^{\mu}
\ar[d]_{\epsilon\otimes \epsilon}
& B
\ar[d]^{\epsilon}
 \\
\mathbf{k}\otimes\mathbf{k}
\ar[r]
& \mathbf{k}
}
}\]
and
\[
\centerline{
\xymatrix{
\mathbf{k}
\ar[dr]_{\eta}
\ar[rr]_{id}
&& \mathbf{k},
\\
& B
\ar[ur]_{\epsilon}
}
}\] where $\tau:a\otimes b\mapsto b\otimes a$ is the \emph{flip map}.

The contents of the diagrams can be expressed as
\begin{eqnarray}
  \Delta(a\cdot b) &=& a_1\cdot b_1\otimes a_2\cdot b_2 \label{p1}\\
  \Delta (\mathbf{1}) &=& \mathbf{1}\otimes \mathbf{1} \label{p2}\\
  \epsilon(a\cdot b) &=& \epsilon(a)\epsilon(b)  \label{p3}\\
  \epsilon(\mathbf{1}) &=& 1 ,\label{p4}
\end{eqnarray}
where $a,b\in B$ and the multiplication in $\mathbf{k}$ is indicated by juxtaposition. These expressions will be very useful.

Let now $A$ be an algebra and $C$ a coalgebra. Take $ \operatorname{Hom}(C,A)$ ($C$ and $A$ regarded as vector spaces). We define for $f,g \in\operatorname{Hom}(C,A)$ the operation of \emph{convolution}
\begin{equation}
    f\ast g=\mu_A(f\otimes g)\Delta_C.
\end{equation}

$ \operatorname{Hom}(C,A)$ has a natural algebra structure with structural maps given by  
\begin{eqnarray}
  \mu_{\operatorname{Hom}(C,A)} (f\otimes g)&=& f\ast g \\
  \eta_{\operatorname{Hom}(C,A)} (\lambda) &=& \lambda\eta_A\circ\epsilon_C,
\end{eqnarray}
$\lambda\in\mathbf{k}$.

Now consider a bialgebra $H$. If there exists an element $S\in \operatorname{Hom}(H,H)$ such that
\begin{equation}\label{antipoda}
    S\ast \mathbf{1}_{\operatorname{Hom}(H,H)}=\mathbf{1}_{\operatorname{Hom}(H,H)}\ast S=\eta_H\circ\epsilon_H,
\end{equation}
$(H, \mu, \eta, \Delta, \epsilon, S)$, or for simplicity $H$, is called a \emph{Hopf algebra}. The element $S:H\rightarrow H$ is called \emph{antipode} or \emph{coinverse}. If this element exists, it is unique, which comes from the fact that it is a left and right inverse.

By applying the definition of convolution, one can see that equation (\ref{antipoda}) is equivalent to
\begin{equation}\label{anti2}
    \mu\circ(S\otimes\ id)\circ\Delta=\mu(id\otimes S)\Delta=\eta\circ\epsilon,
\end{equation}
so that the definition of Hopf algebra can be apprehended by the commutativity of the diagram
\[
\centerline{
\xymatrix{
&H\otimes H
\ar[rr]^{S\otimes id}
&& H\otimes H
\ar[rd]^{\mu}
&
\\
H
\ar[ur]^{\Delta}
\ar[rr]^{\epsilon}
\ar[dr]_{\Delta}
&& \mathbf{k}
\ar[rr]^{\eta}
&& H.
\\
&H\otimes H
\ar[rr]^{id\otimes S}
&& H\otimes H
\ar[ru]_{\mu}
&
}
}\]

As a direct consequence of the definition, $S$ is an antiautomorphism of $H$ and  
\begin{eqnarray}
  S(a\cdot b) &=& S(b)\cdot S(a) \label{118} \\ 
  S(\mathbf{1}) &=& \mathbf{1} \label{s1} \\
  \epsilon(S(a)) &=& \epsilon(a)  \\
  \Delta(S(a)) &=& a_2\otimes a_1 \label{121}
\end{eqnarray}
hold for $a,b\in H$.

If $S^2=\mathbf{1}_{\operatorname{Hom}(H,H)}=id_H$, $H$ is said to be \emph{involutive}. In particular, if $H$ is commutative or cocommutative ($\tau\circ\Delta=\Delta$), then $H$ is involutive.

We now turn our attention to the representations of a Hopf algebra, which will be given by the action of 
 $H$ (regarded simply as an algebra) on a module. 

Let $M$ be a $\mathbf{k}$-module (i.e., a vector space; for more details see chapter \textbf{12} of \cite{artin}) and $H$ an algebra. We say that $M$ is a \emph{left $H$-module} if there exists a linear map $$\alpha: H\otimes M\rightarrow M$$ such that the diagrams  
\[
\centerline{
\xymatrix{
H\otimes H\otimes M
\ar[r]^{id\otimes\alpha}
\ar[d]_{\mu\otimes id}
& H\otimes M
\ar[d]^{\alpha}\\
H\otimes M
\ar[r]^{\alpha}
& M
}
}
\]
and
\[
\centerline{
\xymatrix{
\mathbf{k}\otimes M 
\ar[r]^{\eta\otimes id}
\ar[d]
& H\otimes M
\ar[d]^{\alpha}\\
M
\ar[r]^{id_M}
& M
}
}
\]
commute.

The map $\alpha$ is called a  \emph{left action} of $H$ on $M$ and the pair $(\alpha, M)$ a \emph{representation} of $H$. 

Equivalently, we can say that the map
\begin{eqnarray}
\rho:H&\rightarrow&\operatorname{End}(M)\nonumber\\
h&\mapsto&\alpha(h\otimes -)
\end{eqnarray}
is an algebra homomorphism (with the multiplication in $\operatorname{End}(M)$ given by the composition), which is a more conventional way of seeing that $(\alpha, M)$ is a representation of $H$.

For simplicity, we denote the action by the symbol $\triangleright$, that is, $\alpha(h,v)=h\triangleright v$, $h\in H, v\in M$. The commutativity of the diagrams can thus be expressed as
\begin{eqnarray}
(h\cdot g)\triangleright v&=&h\triangleright(g\triangleright v)\\
\mathbf{1}\triangleright v&=& v,
\end{eqnarray}
$\forall g,h\in H, \forall v\in M$.

All these notions can be dualized to the notions of \emph{$H$-comodule}, \emph{coaction} and \emph{corepresentation}. The ideas of \emph{right $H$-module} and \emph{right action} are also entirely analogous.

Let us now consider the more interesting case where the left $H$-module $M$ is an algebra, with multiplication $m:M\otimes M\rightarrow M$. We say that $H$ acts \emph{covariantly} on $M$ if the multiplication $m$ is respected by the action of $H$, i.e.,
\begin{eqnarray}
h\triangleright(m(v\otimes w))&=&m((h_1\triangleright v)\otimes(h_2\triangleright w))\label{cov1}\\
h\triangleright \mathbf{1}_M&=&\epsilon(h)\mathbf{1}_M\label{cov2},
\end{eqnarray}
$\forall h\in H$, $\forall v,w\in M$. In this case, we say that $m$ is \emph{equivariant} with respect to the action $\alpha$, and that $M$ is an \emph{$H$-module algebra}.

There are some remarkable actions of $H$ on itself, like the so-called \emph{regular action}, given by the multiplication $\mu$, and the \emph{adjoint action}, given by
\begin{equation}\label{ad}
\operatorname{ad}_g(h)=g_1\cdot h\cdot S(g_2),
\end{equation}
$g,h\in H$. The adjoint action makes $H$ an $H$-module algebra and furnishes the \emph{adjoint representation} of $H$.

\section{Drinfel'd Twist}

In this section we shall introduce the notion of quasi-triangular Hopf algebras, first introduced in \cite{drin} (for a review, see \cite{majid2}) and present a systematic method of building quasi-triangular Hopf algebras, the so-called \emph{Drinfel'd twist}.

A Hopf algebra $H$ is \emph{quasi-cocommutative} if there is an invertible element $\mathcal{R}\in H\otimes H$ such that
\begin{equation}\label{qc}
(\tau\circ\Delta)(a)=\mathcal{R}\cdot\Delta(a)\cdot\mathcal{R}^{-1}
\end{equation}
$\forall a\in H$.

Denoting
\begin{equation}
\mathcal{R}=\mathcal{R}^\alpha\otimes\mathcal{R}_\alpha, \quad \mathcal{R}^{-1}=\bar{\cal R}^\alpha\otimes\bar{\cal R}_\alpha, 
\end{equation}
where a sum over the multi-index $\alpha$ is understood, it is convenient to introduce 
\begin{eqnarray}
\mathcal{R}_{12}&=&\mathcal{R}^\alpha\otimes\mathcal{R}_\alpha\otimes\mathbf{1}\\
\mathcal{R}_{13}&=&\mathcal{R}^\alpha\otimes\mathbf{1}\otimes\mathcal{R}_\alpha\\
\mathcal{R}_{23}&=&\mathbf{1}\otimes\mathcal{R}^\alpha\otimes\mathcal{R}_\alpha.
\end{eqnarray}

A quasi-cocommutative Hopf algebra is said to be \emph{quasi-triangular} if
\begin{eqnarray}
(\Delta\otimes id)\mathcal{R}&=&\mathcal{R}_{13}\mathcal{R}_{23}\\
(id\otimes \Delta)\mathcal{R}&=&\mathcal{R}_{13}\mathcal{R}_{12}.\label{ax2}
\end{eqnarray}
The element $\mathcal{R}$ is called \emph{quasi-triangular structure} or \emph{universal R-matrix}. Additionally, if $\mathcal{R}^{-1}=\tau(\mathcal{R})$, $H$ is \emph{triangular}. Every cocommutative Hopf algebra is trivially triangular with $\mathcal{R}=\mathbf{1}\otimes \mathbf{1}$. 

As a consequence of the definition, we have that the quasi-triangular structure $\mathcal{R}$ satisfies 
\begin{equation}\label{qybe}
\mathcal{R}_{12}\mathcal{R}_{13}\mathcal{R}_{23}=\mathcal{R}_{23}\mathcal{R}_{13}\mathcal{R}_{12},
\end{equation}
which is the \emph{quantum Yang-Baxter equation} ([29--31], \cite{baxter}). This equation, for its importance in several physical systems, served as original motivation for the study of quasi-triangular Hopf algebras: for every matricial representation $\rho$ of $H$, $(\rho\otimes\rho)\mathcal{R}$ is a matricial solution of (\ref{qybe}), hence the name \emph{universal R-matrix}. For more details, see e.g. \cite{majid2}.

In order to prove that $\mathcal{R}$ satisfies the quantum Yang-Baxter equation, we apply $(id\otimes\tau)$ to (\ref{ax2}) obtaining
\begin{equation}
(id\otimes\tau\circ\Delta)\mathcal{R}=(id\otimes\tau)\mathcal{R}_{13}\mathcal{R}_{12}=\mathcal{R}_{12}\mathcal{R}_{13},
\end{equation}
and, on the other hand, we make use of the quasi-cocommutativity condition (\ref{qc})
\begin{equation}
(id\otimes\tau\circ\Delta)\mathcal{R}=\mathcal{R}_{23}((id\otimes\Delta)\mathcal{R})\mathcal{R}_{23}^{-1}=\mathcal{R}_{23}\mathcal{R}_{13}\mathcal{R}_{12}\mathcal{R}_{23}^{-1}.
\end{equation}
Since $\mathcal{R}_{23}$ is invertible, (\ref{qybe}) is verified. 

We shall now present the method of the Drinfel'd twist (see \cite{res}), which can be used to produce quasi-triangular Hopf algebras from cocommutative Hopf algebras. Let us begin with some definitions. Let $H$ be a Hopf algebra. A \emph{2-cocycle} is an invertible element $\xi\in H\otimes H$ such that 
\begin{equation}\label{2co}
(\mathbf{1}\otimes\xi)(id\otimes\Delta)\xi=(\xi\otimes\mathbf{1})(\Delta\otimes id)\xi.
\end{equation}
The 2-cocycle $\xi$ is said to be \emph{counital} if
\begin{equation}\label{cou}
(\epsilon\otimes id)\xi=\mathbf{1}=(id\otimes \epsilon)\xi.
\end{equation}

Let now $(H, \mu, \eta, \Delta, \epsilon, S)$ be a cocommutative Hopf algebra and $\mathcal{F}\in H\otimes H$ a counital 2-cocycle. We have that $\chi=\mu(id\otimes S)\mathcal{F}$ is an invertible element of $H$ with $\chi^{-1}=\mu(S\otimes id)\mathcal{F}^{-1}$. 

Defining $\Delta^\mathcal{F}:H\rightarrow H\otimes H$ and $S^\mathcal{F}:H\rightarrow H$ as
\begin{eqnarray}
\Delta^{\mathcal{F}}(a) &=& \mathcal{F}\Delta(a) \mathcal{F}^{-1} \label{copdef} \\
S^\mathcal{F}(a)&=&\chi S(a) \chi^{-1}, \label{antdef}
\end{eqnarray}
$(H, \mu, \eta, \Delta^\mathcal{F}, \epsilon, S^\mathcal{F})$ is a triangular Hopf algebra with universal R-matrix given by 
\begin{equation}\label{r}
\mathcal{R}=\mathcal{F}_{21}\mathcal{F}^{-1}.
\end{equation}

We denote the twisted Hopf algebra $(H, \mu, \eta, \Delta^\mathcal{F}, \epsilon, S^\mathcal{F})$ just $H^\mathcal{F}$. It should be pointed out that, as an algebra, $H^\mathcal{F}$ is identical to $H$ (i.e., they are the same vector space and the algebric structures remain untwisted). The element $\mathcal{F}$ is called \emph{twist}, and the notation
\begin{equation}
\mathcal{F}=f^\alpha\otimes f_\alpha, \quad \mathcal{F}^{-1}=\bar{f}^\alpha\otimes \bar{f}_\alpha,
\end{equation}
with a sum over the multi-index $\alpha$, will be employed often.

To prove that $H^\mathcal{F}$ is a Hopf algebra, we must show that $\Delta^\mathcal{F}$ is coassociative, i.e.,
\begin{equation}
(\Delta^\mathcal{F}\otimes id)\Delta^\mathcal{F}(a)=(id\otimes\Delta^\mathcal{F})\Delta^\mathcal{F}(a),
\end{equation}
and that $S^\mathcal{F}$ is an antipode, that is,  
\begin{equation}
(S^\mathcal{F}\otimes id)\Delta^\mathcal{F}(a)=(id\otimes S^\mathcal{F})\Delta^\mathcal{F}(a),
\end{equation}
$\forall a\in H$. The demonstration involves only direct application of the definitions, of the properties (\ref{118}), (\ref{121}) and of the condition of counital 2-cocycle (\ref{2co}), (\ref{cou}). 

We must now prove the triangularity of $H^\mathcal{F}$. Initially, we verify that $H^\mathcal{F}$ is quasi-cocommutative:
\begin{equation}
\tau\circ\Delta^\mathcal{F}=\mathcal{F}_{21}\Delta\mathcal{F}_{21}^{-1}=\mathcal{R}\mathcal{F}\Delta\mathcal{F}^{-1}\mathcal{R}^{-1}=\mathcal{R}\Delta^\mathcal{F}\mathcal{R}^{-1}.
\end{equation} 
We must also prove that $\mathcal{R}$ defined in (\ref{r}) is a quasi-triangular structure, i.e.,
\begin{eqnarray}
(\Delta^\mathcal{F}\otimes id)\mathcal{R}&=&\mathcal{R}_{13}\mathcal{R}_{23}\\
(id\otimes \Delta^\mathcal{F})\mathcal{R}&=&\mathcal{R}_{13}\mathcal{R}_{12},
\end{eqnarray}
which requires only manipulations of the basic properties and of the 2-cocycle condition, as well as the cocommutativity of  $H$. Finally, it is evident that 
\begin{equation}
\mathcal{R}_{21}=\mathcal{F}\mathcal{F}_{21}^{-1}=\mathcal{R}^{-1}.
\end{equation}

Now take an $H$-module algebra $M$ with multiplication $m$ on which $H$ acts covariantly in the sense of (\ref{cov1})--(\ref{cov2}). By definition, 
\begin{equation}\label{multdef}
v\star w=m^\mathcal{F}(v\otimes w)=m(\mathcal{F}^{-1}\triangleright (v\otimes w)),
\end{equation}
$v,w\in M$. In this case, $m^\mathcal{F}$ defines a new associative algebra $M_\star$ which is covariant under the action of the twisted Hopf algebra $H^\mathcal{F}$ defined above. The proof of this statement is simple. The associativity of $\star$ follows from the 2-cocycle condition satisfied by $\mathcal{F}$. To prove the equivariance of $m^\mathcal{F}$, it suffices to see that $h\triangleright(m^\mathcal{F}(v\otimes w))=m^\mathcal{F}(\Delta^\mathcal{F}(h)\triangleright(v\otimes w))$: 
\begin{eqnarray}
h\triangleright(m^\mathcal{F}(v\otimes w))=h\triangleright m(\mathcal{F}^{-1}\triangleright(v\otimes w)) &=& m(\Delta(h)\mathcal{F}^{-1}\triangleright(v\otimes w))=\nonumber \\
= m(\mathcal{F}^{-1}\mathcal{F}\Delta(h)\mathcal{F}^{-1}\triangleright(v\otimes w))&=&m^\mathcal{F}(\Delta^\mathcal{F}(h)\triangleright(v\otimes w)).
\end{eqnarray}

\section{Universal Enveloping Algebra and its Drinfel'd Twist}

In this work, the only Hopf algebra which we will use is the \emph{universal enveloping algebra} of a Lie algebra over the complex number field $\mathbb{C}$.

Let us first recall the definition of a Lie algebra. Let $\mathfrak{g}$ be a vector space over the field $\mathbf{k}$ and $[\cdot,\cdot]:\mathfrak{g}\times \mathfrak{g}\rightarrow \mathfrak{g}$ a bilinear binary operation ($ \left[ \alpha x+ \beta y, z \right] = \alpha\left[x,z\right]+\beta\left[y,z\right]$ e $\left[z, \alpha x+\beta y\right] =  \alpha\left[z,x\right]+\beta\left[z,y\right]$  $\forall \alpha,\beta\in\mathbf{k}$, $x,y,z\in\mathfrak{g}$).

If $[\cdot,\cdot]$ satisfies the property of skew-symmetry and the Jacobi identities
\begin{eqnarray}
  \left[x,y\right]+\left[y,x\right] &=& 0 \\
  \left[x,\left[y,z\right]\right]+\left[z,\left[x,y\right]\right]+\left[y,\left[z,x\right]\right] &=& 0,
\end{eqnarray}
$(\mathfrak{g}, [\cdot,\cdot])$, or simply $\mathfrak{g}$, is called a \emph{Lie algebra}.

Lie algebras are not associative, which makes the direct application of the Drinfel'd twist impossible. Therefore, it is necessary to find a unital associative algebra which contains the vector space $\mathfrak{g}$. The natural construction to make is the universal enveloping algebra of $\mathfrak{g}$, which has the important properties of having $\mathfrak{g}$ as a linear subspace and exhibiting a natural Hopf algebra structure.

Let $\mathfrak{g}$ be a Lie algebra over $\mathbf{k}$. Consider the tensor algebra of $\mathfrak{g}$
\begin{equation}
    T(\mathfrak{g})=\bigotimes\!^\bullet \mathfrak{g}=\mathbf{k}\oplus\mathfrak{g}\oplus(\mathfrak{g}\otimes\mathfrak{g})\oplus(\mathfrak{g}\otimes\mathfrak{g}\otimes\mathfrak{g})\oplus\ldots
\end{equation}

Let now $I$ be the (two-sided) ideal in $T(\mathfrak{g})$ generated by the set of all elements of the form $(x\otimes y - y \otimes x)-[x,y]$. The \emph{universal enveloping algebra} of $\mathfrak{g}$ is defined as the quotient
\begin{equation}
    \mathcal{U}(\mathfrak{g})=T(\mathfrak{g})/I.
\end{equation}

The Poincar\'e-Birkhoff-Witt theorem provides a more explicit, tractable description of the universal enveloping algebra $\mathcal{U}(\mathfrak{g})$, as well as some important consequences. (For further details, see chapter \textbf{5} in \cite{jacobson}.)

Let $\mathfrak{g}$ be a Lie algebra, $\mathcal{U}(\mathfrak{g})$ its universal enveloping algebra and  $\{\tau_i\}$ a \emph{totally ordered basis} of $\mathfrak{g}$ with elements satisfying the commutation relations $[\tau_i,\tau_j]=iC_{ij}^k\tau_k$. The affirmation of the Poincar\'e-Birkhoff-Witt theorem is that the set of monomials $\{\tau_{i_1}\cdots\tau_{i_n}\}$ with $i_1\leq\cdots\leq i_n$ is a basis for $\mathcal{U}(\mathfrak{g})$.

This result gives us a much more convenient description of $\mathcal{U}(\mathfrak{g})$: that $\mathcal{U}(\mathfrak{g})$ is the algebra of the polynomials of the generators  $\tau_i$ modulo the commutation relations $[\tau_i,\tau_j]=iC_{ij}^k\tau_k$, where $C_{ij}^k$ are the  \emph{structure constants} of the Lie algebra. We shall adopt this point of view in the following chapters.

As a corollary of the theorem, we have that the canonical homomorphism $\iota:\mathfrak{g}\rightarrow\mathcal{U}(\mathfrak{g})$ is an injection and that $\mathfrak{g}$ is, in particular, the linear subspace of $\mathcal{U}(\mathfrak{g})$.

The universal enveloping algebra of a Lie algebra has a natural Hopf algebra structure, inherited from its tensor algebra structure, with coalgebra maps given by
\begin{eqnarray}
  \Delta(\tau_i) &=& \tau_i\otimes \mathbf{1} + \mathbf{1}\otimes \tau_i\label{u1}\\
  \epsilon (\tau_i) &=& 0,\label{u2}
\end{eqnarray}
and antipode given by
\begin{equation}\label{u3}
    S(\tau_i)=-\tau_i.
\end{equation}

To prove that (\ref{u1})--(\ref{u3}) furnish a  Hopf algebra structure, it suffices to verify properties (\ref{p1}) and (\ref{p3}):
\begin{eqnarray}
    \Delta(\tau_i\cdot\tau_j)&=&(\tau_i\otimes \mathbf{1} + \mathbf{1}\otimes \tau_i)\cdot(\tau_j\otimes \mathbf{1} + \mathbf{1}\otimes \tau_j)\nonumber\\
    &=&\tau_i\cdot\tau_j\otimes \mathbf{1} + \tau_i\otimes \tau_j+\tau_j\otimes \tau_i + \mathbf{1}\otimes \tau_i\cdot\tau_j \nonumber\\
    &=& (\tau_i)_1\cdot(\tau_j)_1\otimes(\tau_i)_2\cdot(\tau_j)_2,
\end{eqnarray}
where it is important to stress the double application of the Sweedler notation and, therefore, the occurrence of a double sum;
\begin{eqnarray}
    \epsilon(\tau_i\cdot\tau_j)=\epsilon(\tau_i)\epsilon(\tau_j)=0,
\end{eqnarray}
and, finally, show that $S$ is an antipode, which can be done by applying the explicit definition given in (\ref{anti2}):
\begin{eqnarray}
    \mu(S\otimes id)\Delta(\tau_i) &=& \mu(S(\tau_i)\otimes\mathbf{1}+S(\mathbf{1})\otimes\tau_i)=-\tau_i+\tau_i= \nonumber\\
    \mu(id\otimes S)\Delta(\tau_i) &=& \mu(\tau_i\otimes S(\mathbf{1})+\mathbf{1}\otimes S(\tau_i) )=\tau_i-\tau_i=\nonumber\\
    &=& 0 = \eta(\epsilon(\tau_i)) = \epsilon(\tau_i)\mathbf{1}.
\end{eqnarray}

By means of the multiplicativity and linearity of $\Delta$ and $\epsilon$ and of the antimultiplicativity of $S$, these definitions can be extended to all monomials of $\tau_i$ and thus to all elements of $\mathcal{U}(\mathfrak{g})$. With (\ref{p2}), (\ref{p4}) and (\ref{s1}), the description of $\mathcal{U}(\mathfrak{g})$ as a Hopf algebra is complete. 

For the universal enveloping algebra $\mathcal{U}(\mathfrak{g})$, the adjoint action (\ref{ad}) is the Lie commutator, as can be directly verified:
\begin{equation}
\operatorname{ad}_{\tau_i}(\tau_j)=(\tau_i)_1\cdot \tau_j\cdot S\left((\tau_i)_2\right)=\tau_i\cdot \tau_j\cdot \mathbf{1}+\mathbf{1}\cdot \tau_j\cdot (-\tau_i)=[\tau_i,\tau_j].
\end{equation}

We can now apply the prescription presented in the previous section to the universal enveloping algebra  $\mathcal{U}(\mathfrak{g})$. Take an element
\begin{equation}
\mathcal{F}\in\mathcal{U}(\mathfrak{g})\otimes\mathcal{U}(\mathfrak{g})
\end{equation}
which is a  twist, i.e., satisfies the conditions (\ref{2co}) and (\ref{cou}). By using the expressions (\ref{copdef}) and (\ref{antdef}), we calculate the deformed coproduct and antipode of the primitive elements $\tau_i\in\mathfrak{g}$
\begin{eqnarray}
\Delta^{\mathcal{F}}(\tau_i) &=& \mathcal{F}\Delta(\tau_i) \mathcal{F}^{-1} \\
S^\mathcal{F}(\tau_i)&=&\chi S(\tau_i) \chi^{-1}, 
\end{eqnarray}
with the algebric structures and the counit remaining unchanged. We call the twisted Hopf algebra thus obtained $\mathcal{U}^{\cal F}(\mathfrak{g})$. 

It is clear that $\mathfrak{g}$ is not the linear subspace of $\mathcal{U}^{\cal F}(\mathfrak{g})$. It is only natural, therefore, to investigate which is the linear subspace of $\mathcal{U}^{\cal F}(\mathfrak{g})$. We will call this space $\mathfrak{g}^{\cal F}$ and its elements $\tau_i^{\cal F}$ \emph{deformed generators}. 

The conditions for $\mathfrak{g}^{\cal F}$, pointed out in \cite{wor}, are three, namely: (i) that $\{\tau_i^{\cal F}\}$ is a basis of $\mathfrak{g}^{\cal F}$, (ii) the minimal deformation of the Leibniz rule
\begin{equation}
\Delta^{\cal F}(\tau_i^{\cal F})=\tau_i^{\cal F}\otimes\mathbf{1}+f_i^j\otimes\tau_j^{\cal F},
\end{equation}
$f_i^j\in\mathcal{U}(\mathfrak{g})$, and (iii) that, under deformed adjoint action, denoted  $[\cdot,\cdot]_{\cal F}$, the structure constants of $\mathfrak{g}$ are reproduced
 \begin{equation}
[\tau_i^{\cal F},\tau_j^{\cal F}]_\mathcal{F}=iC_{ij}^k\tau_k^{\cal F}.
\end{equation}

To obtain $\mathfrak{g}^{\cal F}$, there is a canonical procedure (see \cite{aschieri}, \cite{aschieri2}). Take as deformed generators
 \begin{equation}\label{gerdef}
\tau_i^{\cal F}=\bar{f}^\alpha(\tau_i)\bar{f}^\alpha
\end{equation}
with deformed coproduct given by
 \begin{equation}\label{cdd}
\Delta^{\cal F}(\tau_i^{\cal F})=\tau_i^{\cal F}\otimes\mathbf{1}+\bar{\cal R}^\alpha\otimes\bar{\cal R}_\alpha(\tau_i^{\cal F}).
\end{equation}

In accordance with (\ref{ad}), the deformed adjoint action is given by
 \begin{equation}\label{addef}
[\tau_i^{\cal F},\tau_j^{\cal F}]_\mathcal{F}=(\tau_i^{\cal F})_1\cdot \tau_j^{\cal F}\cdot S^{\cal F}\left((\tau_i^{\cal F})_2\right).
\end{equation}

Constructed in this way, $\mathfrak{g}^{\cal F}$ meets the three requirements above.

Finally, since we intend to study fermionic systems, we will make use of the so-called  \emph{Lie superalgebras} (or $\mathbb{Z}_2$-graded Lie algebras) and their twisted Hopf algebras.

A \emph{Lie superalgebra} over a field $\mathbf{k}$ (of characteristic zero) is a vector space $\mathfrak{g}=\mathfrak{g}_0\oplus\mathfrak{g}_1$ endowed with a  bilinear binary operation $[\cdot,\cdot\}:\mathfrak{g}\times\mathfrak{g}\rightarrow\mathfrak{g}$ satisfying  the properties of $\mathbb{Z}_2$-grading
\begin{equation}
[\mathfrak{g}_i,\mathfrak{g}_j\}\subset\mathfrak{g}_{(i+j\!\!\!\!\mod 2)},
\end{equation}
 $\mathbb{Z}_2$-graded skew-symmetry
\begin{equation}
[x,y\}=(-1)^{|x||y|}[y,x\},
\end{equation}
and generalized Jacobi identities
\begin{equation}
(-1)^{|x||z|}[[x,y\},z\}+(-1)^{|y||x|}[[y,z\},x\}+(-1)^{|z||y|}[[z,x\},y\}=0,
\end{equation}
$\forall x,y,z\in\mathfrak{g}$, where $|x|=i$ if $x\in\mathfrak{g}_i$ (\emph{degree} of $x$).  We call $\mathfrak{g}_0$ the \emph{even} or \emph{bosonic} part of $\mathfrak{g}$ and $\mathfrak{g}_1$ the \emph{odd} or \emph{fermionic} part of $\mathfrak{g}$. The Lie superalgebra $\mathfrak{g}$ can also be extended to its \emph{universal enveloping superalgebra}, with a Hopf algebra strucure.

Formulas (\ref{118}), (\ref{r}), (\ref{gerdef}) and (\ref{addef}) are naturally extended to the case of  superalgebras as
  \begin{eqnarray}
    S(\tau_i\cdot\tau_j)&=&(-1)^{|\tau_i||\tau_j|}S(\tau_j)\cdot S(\tau_i) \\
  \mathcal{R}&=&\sum_{\alpha,\beta}(-1)^{|\bar{f}^\beta||f^\alpha|}(f_\alpha\bar{f}^\beta\otimes f^\alpha\bar{f}_\beta)\\
    \tau_i^{\mathcal{F}}&=&\sum_\alpha(-1)^{|\bar{f}_\alpha||\tau_i|}\bar{f}^\alpha(\tau_i)\bar{f}_\alpha \\
\left[\tau_i^\mathcal{F},\tau_j^\mathcal{F}\right\}_\mathcal{F}&=&\sum_k(-1)^{|\tau_j^\mathcal{F}||(\tau_i^\mathcal{F})^k_2|}(\tau_i^\mathcal{F})^k_1\cdot\tau_j^\mathcal{F}\cdot S^{\cal F}\left((\tau_i^\mathcal{F})^k_2\right).
\end{eqnarray}

\chapter{Bosonic Heisenberg Algebras: First and Second Quantization}

In this chapter we will show that (contrarily to the stated in \cite{barut} and \cite{palev})  it is indeed possible to construct the universal enveloping algebra of the Heisenberg algebra and deform it by means of a Drinfel'd twist, obtaining, as a result, a noncommutative theory. We will also show how this structure can be recovered in the context of second quantization by using the formalism of Wigner oscillators.

\section{Twisted Heisenberg Algebra}

In this section, we apply the formalism outlined in the previous chapter to the bosonic Heisenberg algebra, which we shall denote $\mathfrak{h}(\mathcal{N})$. 
 
Consider the algebra $\mathfrak{h}(\mathcal{N})$:
\begin{eqnarray}\label{algebra1}
   \left[x_i,x_j\right]&=&\left[p_i,p_j\right]=0 \label{ha} \\
  \left[x_i,p_j\right]&=&i\delta_{ij}{\hbar} \\
   \left[{\hbar},x_i\right]&=&\left[{\hbar},p_j\right]=0,\label{algebra2}
\end{eqnarray}
$i,j=1,...,\mathcal{N}$.

We will apply the twist
\begin{equation}\label{twist}
    \mathcal{F}=\exp\left({\frac{i}{2}\alpha_{ij}p_i\otimes
    p_j}\right),
\end{equation}
where $\alpha_{ij}$ is a skew-symmetric matrix.  The 2-cocycle condition (\ref{2co}) is trivially satisfied because $\mathcal{F}$ is an  \emph{Abelian} twist, i.e., it involves only generators that commute among themselves.

We should now stress the necessity of correctly identifying the role of the central extension  $\hbar$ as an element of the algebra $\mathfrak{h}(\mathcal{N})$, and not as a multiple of the identity. It must therefore be treated in an identical manner as the other generators of $\mathfrak{h}(\mathcal{N})$, i.e., with coproduct and antipode given by
\begin{eqnarray}
\Delta({\hbar})&=&{\hbar}\otimes\mathbf{1}+\mathbf{1}\otimes {\hbar} \\
S({\hbar})&=&-{\hbar}.
\end{eqnarray}

We can now proceed to the twisting of $\mathfrak{h}(\mathcal{N})$.

With the help of the Hadamard formula, we calculate the deformed coproduct of $x_k$:
\begin{eqnarray}
\nonumber\Delta^\mathcal{F}(x_k)&=&\exp\left(\frac{i}{2}\alpha_{ij}p_i\otimes p_j\right)\Delta(x_k)\exp\left(-\frac{i}{2}\alpha_{ij}p_i\otimes p_j\right) = \\
    &=&x_k\otimes\mathbf{1}+\mathbf{1}\otimes x_k+\frac{\alpha_{kj}}{2}\left({\hbar}\otimes p_j-p_j\otimes{\hbar}\right).
\end{eqnarray}

Since $p_i$ and ${\hbar}$ commute with the $p_j$s of the twist, their coproducts do not get deformed:
\begin{eqnarray}
    \Delta^\mathcal{F}(p_k)&=&\Delta(p_k) \\
\Delta^\mathcal{F}({\hbar})&=&\Delta({\hbar}).
\end{eqnarray}

The antipode remains undeformed. Seeing this amounts to calculating the element
\begin{equation}
    \chi\equiv f^{\alpha}S(f_\alpha)=\exp\left(-\frac{i}{2}{\alpha_{ij}p_ip_j}\right)=\mathbf{1},
\end{equation}
so that $S^\mathcal{F}=\chi S\chi^{-1}=S.$

The deformation of $x_k$ is given by (\ref{gerdef}):
\begin{equation}\label{def}
    x_k^\mathcal{F}=\bar{f}^\alpha(x_k)\bar{f}_\alpha=x_k-\frac{\alpha_{kj}}{2}{\hbar}
    p_j,
\end{equation}
whereas $p_i$ and ${\hbar}$, for the reason above, undergo no deformation.

We can now calculate the deformed coproducts of the deformed generators. In this instance, the universal R-matrix is simply  $\mathcal{R}=\mathcal{F}^{-2}$ (because $\mathcal{F}_{21}=\mathcal{F}^{-1}$ by the skew-symmetry of $\alpha_{ij}$ and the abelianity of the twist). Thus, the deformed coproduct of $x_k^\mathcal{F}$ is obtained by using (\ref{cdd}):
\begin{equation}
    \Delta^\mathcal{F}(x_k^\mathcal{F})=x_k^\mathcal{F}\otimes\mathbf{1}+\mathbf{1}\otimes
    x_k^\mathcal{F}+\alpha_{ik}p_i\otimes{\hbar},
\end{equation}
where the contribution of ${\hbar}$ should be stressed.

The antipode of  $x_k^\mathcal{F}$ is easily obtained by the antimultiplicative property of $S$: 
\begin{equation}
    S(x_k^\mathcal{F})=-x_k-\frac{1}{2}\alpha_{kj}{\hbar}
    p_j=-x_k^\mathcal{F}-\alpha_{kj}{\hbar}
    p_j.
\end{equation}

We are now in a position of working out the deformed brackets of the deformed generators, according to (\ref{addef}):
\begin{eqnarray}
  \left[x_i^\mathcal{F},p_j^\mathcal{F}\right]_\mathcal{F}&=&i\delta_{ij}{\hbar}\\
  \left[x_i^\mathcal{F},x_j^\mathcal{F}\right]_\mathcal{F}&=&0 \\
  \left[p_i^\mathcal{F},p_j^\mathcal{F}\right]_\mathcal{F}&=&0 \\
  \left[{\hbar}^\mathcal{F},x_i^\mathcal{F}\right]_\mathcal{F}&=&\left[{\hbar}^\mathcal{F},p_i^\mathcal{F}\right]_\mathcal{F}=0.
\end{eqnarray}

It should be noted that the deformed brackets of the deformed quantities yield the same structure constants as the ordinary brackets of the undeformed quantities. The same is observed in \cite{aschieri3} for the universal enveloping algebra of the  Poincar\'e algebra, $\mathcal{U}(iso(1,3))$.

We now calculate the ordinary brackets of the deformed quantities. They are
\begin{eqnarray}
  \left[x_i^\mathcal{F},p_j^\mathcal{F}\right]&=&i\delta_{ij}{\hbar}\\
  \left[x_i^\mathcal{F},x_j^\mathcal{F}\right]&=&i\alpha_{ij}\hbar^2 \label{oc} \\
  \left[p_i^\mathcal{F},p_j^\mathcal{F}\right]&=&0 \\
  \left[{\hbar}^\mathcal{F},x_i^\mathcal{F}\right]&=&\left[{\hbar}^\mathcal{F},p_i^\mathcal{F}\right]=0.
\end{eqnarray}

Notice that the $x_i^\mathcal{F}$s
are of noncommutative nature, as opposed to the original $x_i$s of (\ref{ha}). Commutativity can be restored by the inverse of (\ref{def}), i.e., $x_i=x_i^\mathcal{F}+\frac{\alpha_{ij}}{2}\hbar p_j$, which is known in the literature as \emph{Bopp shift} \cite{bopp}, while the deformation in the costructures cannot be removed by any transformation, and is, in this sense, more fundamental. It is also interesting to recall that $\mathfrak{h}(\mathcal{N})$ is the linear subspace of $\mathcal{U}(\mathfrak{h}(\mathcal{N}))$ and that, analogously, the $x_i^\mathcal{F}$s (along with $p_i^\mathcal{F}=p_i$ and $\hbar^\mathcal{F}=\hbar$) form the linear subspace of $\mathcal{U}^{\mathcal{F}}(\mathfrak{h}(\mathcal{N}))$, which we will call $\mathfrak{h}^{\mathcal{F}}(\mathcal{N})$. 

Finally, we want to study the deformation of the multiplication on the  $\mathcal{U}(\mathfrak{h}(\mathcal{N}))$-module $M$ consisting of the space of functions on $\mathbb{R}^{\cal N}$ endowed with the usual pointwise multiplication
\begin{eqnarray}
m(g\otimes h)&=&g\cdot h,\nonumber\\
(g\cdot h)(\check{x})&=&g(\check{x})h(\check{x}),
\end{eqnarray}
$\check{x}\in\mathbb{R}^\mathcal{N}$.

The action of the algebra $\mathfrak{h}(\mathcal{N})$ on the module is  such that $p_i$ acts by differentiation and   $x_i$ acts by multiplication, 
\begin{eqnarray}
p_i\triangleright g(\check{x})&=&-i\hbar\frac{\partial}{\partial \check{x}_i}g(\check{x})\\
x_i\triangleright g(\check{x})&=&\check{x}_i\cdot g(\check{x}).
\end{eqnarray}

The deformed multiplication on the module (see \cite{asc}) is given by
\begin{eqnarray}
  \nonumber  g(\check{x})\star h(\check{x})&\equiv& m^\mathcal{F}(g(\check{x})\otimes h(\check{x}))=(m\circ\mathcal{F}^{-1})(g(\check{x})\otimes    h(\check{x}))= \\
  \nonumber  &=&\left(\bar{f}^\alpha\triangleright g(\check{x})\right)\cdot\left(\bar{f}_\alpha\triangleright h(\check{x})\right)= \\
   &=& e^{\frac{i}{2}\theta_{ij}\frac{\partial}{\partial\check{x}_i}\frac{\partial}{\partial\check{y}_j}}\left(g(\check{x})\cdot h(\check{y})\right)|_{\check{x}=\check{y}}, 
\end{eqnarray}
where we introduce, for convenience, $\theta_{ij}=\alpha_{ij}\hbar^2$.

The star product in this particular case is commonly known as Weyl-Groenewold product \cite{gro} or Moyal product \cite{moyal}.

Defining the Moyal bracket as $\left[g,h\right]_\star\equiv
(g\star h - h\star g$), we see that noncommutativity among position variables is implemented:
\begin{equation}
 \left[\check{x}_i,\check{x}_j\right]_\star=i\theta_{ij}. 
\end{equation}

Here it was necessary to introduce the notation $\check{x}_i$ for the quantities on the module corresponding to the operators $x_i$. Correspondence between functions of operators of the Heisenberg algebra and functions on the commutative space is given by the Wigner transformation, introduced in \cite{wigner1}. Conversely, to obtain functions of operators of the Heisenberg algebra from functions on the phase space we must use the well-known transformation introduced by Weyl in \cite{weyl}.

\section{Second  Quantization and Wigner Oscillators}

Having studied the twisting of the bosonic Heisenberg algebra, we now intend to show how this structure can be recovered in the context of second quantization, about which we will make a brief digression. In this section, $\hbar$ is a c-number.

Consider the Lagrangian 
\begin{equation}
    L=\int d^Dx \left(\frac{i\hbar}{2}\psi^\ast\stackrel{\leftrightarrow}{\partial_o}\psi-\frac{\hbar^2}{2m}|\vec{\nabla}\psi|^2 \right).
\end{equation}

It is straightforward to show that it has the Schr\"odinger equation
\begin{equation}
    i\hbar\frac{\partial \psi}{\partial t}=-\frac{\hbar^2}{2m}\nabla^2\psi.
\end{equation}
as the equation of motion.

Going to the Hamiltonian formalism, the definition of canonically conjugate momenta gives rise to the constraints
\begin{eqnarray}
\pi_\psi-\frac{i\hbar}{2}\psi^\ast\approx 0 \\
\pi_\psi^\ast-\frac{i\hbar}{2}\psi\approx 0,
\end{eqnarray}
where $\pi_\psi, \pi_\psi^\ast$  are the canonically conjugate momenta to $\psi, \psi^\ast$.

Applying the formalism of Dirac brackets (see \cite{dirac}), one can impose the strong equality on the constraints, obtaining the brackets
\begin{equation}
    \{\psi(\vec{x},t),\psi^\ast(\vec{y},t)\}_{DB}=\frac{1}{i\hbar}\delta^D(\vec{x}-\vec{y}).
\end{equation}

The usual procedure of canonical quantization can now be applied and yields the commutation relations 
\begin{eqnarray}
  \left[ \psi(\vec{x}),\psi^\dag(\vec{y})\right] &=& \delta^D(\vec{x}-\vec{y}) \label{psi1}\\
  \left[ \psi(\vec{x}),\psi(\vec{y})\right] &=& \left[
  \psi^\dag(\vec{x}),\psi^\dag(\vec{y})\right]=0. \label{psi2}
\end{eqnarray}

We define the objects
\begin{eqnarray}
  X_i &=& \int d^Dy\; y_i \psi^\dag(\vec{y})\psi(\vec{y}) \label{xi}\\
  P_i &=& -\frac{i\hbar}{2} \int d^Dy \,
  \psi^\dag(\vec{y})\stackrel{\leftrightarrow}{\partial_i}\psi(\vec{y})
  \label{P}
\end{eqnarray}
which satisfy the commutation relations
\begin{eqnarray}
\left[ X_i,P_j\right]&=&i\hbar\delta_{ij}N\label{xp} \\
\left[ X_i,X_j\right]&=&0 \\
\left[ P_i,P_j\right]&=&0,\label{pp}\\
\left[N, X_i\right]&=&\left[N, P_i\right]=0
\end{eqnarray}
where  $N=\int d^Dy\; \psi^\dag(\vec{y})\psi(\vec{y})$ is the number operator. The expressions (\ref{xi})-(\ref{P}) relate the formalisms of first and second quantization, as will become clear.

The expressions (\ref{xp})-(\ref{pp}) reproduce the Heisenberg algebra, with $\hbar N$ in the role of central extension in the case of $N$ particles. To show that this identification indeed makes sense, we must check that  $X_i$ is a position operator and $P_i$ a momentum operator. Defining $|\vec{y}\rangle={\psi}^\dag(\vec{y})|0\rangle$ and applying (\ref{psi1})-(\ref{psi2}), we obtain, as desired,
\begin{equation}
    X_i|\vec{y}\rangle=y_i|\vec{y}\rangle.
\end{equation}
Straightforward calculation shows that
\begin{eqnarray}
  \left[ P_i,\psi(\vec{x})\right]&=& i\hbar\partial_i\psi(\vec{x}) \\
  \left[ P_i,\psi^\dag(\vec{x})\right]&=& i\hbar\partial_i\psi^\dag(\vec{x}),
\end{eqnarray}
so that $P_i$ is indeed the generator of translations. We have therefore a description of the Heisenberg algebra in the formalism of second quantization.

The fields can now be expanded in their Fourier modes
\begin{eqnarray}
  \psi(\vec{x}) &=& \frac{1}{(2\pi\hbar)^D}\int d^Dp\;e^{\frac{i}{\hbar}\vec{p}\cdot\vec{x}}a_{\vec{p}} \\
  \psi^\dag(\vec{x}) &=& \frac{1}{(2\pi\hbar)^D}\int d^Dp\;e^{-\frac{i}{\hbar}\vec{p}\cdot\vec{x}}a^\dag_{\vec{p}},
\end{eqnarray}
and, conversely,
\begin{eqnarray}
  a_{\vec{p}} &=& \int d^Dx\;e^{-\frac{i}{\hbar}\vec{p}\cdot\vec{x}}\psi(\vec{x})  \\
  a^\dag_{\vec{p}}&=& \int d^Dx\;e^{\frac{i}{\hbar}\vec{p}\cdot\vec{x}}\psi^\dag(\vec{x}) .
\end{eqnarray}

The algebra satisfied by $a_{\vec{p}}$, $a^\dag_{\vec{p}}$ is
\begin{eqnarray}
  \left[ a_{\vec{p}},a^\dag_{\vec{p}'}\right] &=& (2\pi\hbar)^D\delta^D(\vec{p}-\vec{p}') \label{o1} \\
  \left[ a_{\vec{p}},a_{\vec{p}'}\right] &=& \left[ a^\dag_{\vec{p}},a^\dag_{\vec{p}'}\right]=0. \label{o2}
\end{eqnarray}

It is interesting to notice that the algebra of the fields (\ref{psi1})-(\ref{psi2}) and of the corresponding oscillators (\ref{o1})-(\ref{o2}) is the Heisenberg algebra $\mathfrak{h}(\mathcal{N})$ in the ${\cal N}\rightarrow\infty$ limit, with  $\psi^\dag(\vec{x})$ and $a^\dag_{\vec{p}}$ as conjugate momenta to
 $\psi(\vec{x})$ and $a_{\vec{p}}$, respectively, and $\delta^D(\vec{x}-\vec{y})$ and $(2\pi\hbar)^D\delta^D(\vec{p}-\vec{p}')$ in a similar role as $i\hbar\delta_{ij}$. 

We will now try to construct the Hopf algebra structure of the algebras of the fields
 $\psi(\vec{x})$, $\psi^\dag(\vec{x})$ and oscillators  $a_{\vec{p}}$, $a^\dag_{\vec{p}}$, so as to try to deform them by the same twist  (\ref{twist}).

We begin by expressing $\vec{P}$ in momentum space 
\begin{equation}
    P_i=\frac{1}{(2\pi\hbar)^D}\int d^Dp\;p_i a^\dag_{\vec{p}}a_{\vec{p}},
\end{equation}
and obtaining the algebra of the oscillators with $P_i$
\begin{eqnarray}
  \left[ P_i,a_{\vec{p}}\right] &=& -p_ia_{\vec{p}} \\
  \left[ P_i,a^\dag_{\vec{p}}\right] &=& p_ia^\dag_{\vec{p}}.
\end{eqnarray}

We now apply the twist
\begin{equation}
    \mathcal{F}=\exp\left({\frac{i}{2}\alpha_{ij}P_i\otimes
    P_j}\right),
\end{equation}
and obtain the deformation of $a_{\vec{p}}$ and $a^\dag_{\vec{p}}$
\begin{eqnarray}
  a_{\vec{p}}^\mathcal{F} &=& \bar{f}^\alpha(a_{\vec{p}})\bar{f}_\alpha = a_{\vec{p}}\;e^{\frac{i}{2}\alpha_{ij}p_iP_j} \label{a1} \\
  a^{\dag\mathcal{F}}_{\vec{p}} &=&\bar{f}^\alpha(a^\dag_{\vec{p}})\bar{f}_\alpha = a^\dag_{\vec{p}}\;e^{-\frac{i}{2}\alpha_{ij}p_iP_j}, \label{a2}
\end{eqnarray}
which coincides with the findings of \cite{bal2}, as well as the deformation of the fields $\psi(\vec{x})$ and $\psi^\dag(\vec{x})$
\begin{eqnarray}
  \psi^\mathcal{F} (\vec{x})&=& \psi(\vec{x})\;e^{\frac{\hbar}{2}\alpha_{ij}\stackrel{\leftarrow}{\partial_i}P_j}  \label{de1}\\
  \psi^{\dag\mathcal{F}}(\vec{x}) &=& \psi^\dag(\vec{x})\;e^{\frac{\hbar}{2}\alpha_{ij}\stackrel{\leftarrow}{\partial_i}P_j}\label{de2}.
\end{eqnarray}

The deformation of $\psi(\vec{x})$ and $\psi^\dag(\vec{x})$ is compatible with the deformation of $a_{\vec{p}}$ and $a^\dag_{\vec{p}}$, since they are related by the usual Fourier transform 
\begin{eqnarray}
  a_{\vec{p}}^\mathcal{F} &=& \int d^Dx\;e^{-\frac{i}{\hbar}\vec{p}\cdot\vec{x}}\psi^\mathcal{F}(\vec{x}) \label{c1}  \\
  a^{\dag\mathcal{F}}_{\vec{p}}&=& \int d^Dx\;e^{\frac{i}{\hbar}\vec{p}\cdot\vec{x}}\psi^{\dag\mathcal{F}}(\vec{x}) \label{c2} .
\end{eqnarray}

As we have seen, the algebric structure of the Heisenberg algebra is correctly reproduced by the algebras of $\psi(\vec{x})$, $\psi^\dag(\vec{x})$ and $a_{\vec{p}}$, $a^\dag_{\vec{p}}$ (in the sense of (\ref{xp})--(\ref{pp}) with $\hbar N\rightarrow \hat{\hbar}$,  $\hat{\hbar}$ being the central extension of the previous section). We will now try to reproduce the coalgebric structure of the Heisenberg algebra, and will see that in a naive approach the process fails already in the undeformed case.   

The natural construction would be to take the expression (\ref{xi}) and to apply the property of the multiplicativity of the coproduct:
\begin{eqnarray}
  \nonumber \Delta(X_i)&=&\int d^Dy\; y_i \Delta(\psi^\dag(\vec{y})) \Delta(\psi(\vec{y}))=\\
  \nonumber &=&\int d^Dy\; y_i (\psi^\dag(\vec{y})\otimes\mathbf{1}+\mathbf{1}\otimes\psi^\dag(\vec{y}))(\psi(\vec{y})\otimes\mathbf{1}+\mathbf{1}\otimes\psi(\vec{y}))= \\
  &=& X_i\otimes\mathbf{1}+\mathbf{1}\otimes X_i +\int d^Dy\;
  y_i(\psi^\dag(\vec{y})\otimes\psi(\vec{y})+\psi(\vec{y})\otimes\psi^\dag(\vec{y})). \nonumber\\
  \;\label{cope}
\end{eqnarray}

The presence of the cross terms is completely undesirable, because the expected coproduct would be  $\Delta(X_i)=X_i\otimes\mathbf{1}+\mathbf{1}\otimes X_i$. We will show now how to solve this problem.

The solution lies on the notion of \emph{Wigner oscillators}. In \cite{wigner}, Wigner showed that the Heisenberg equations of motion for the position and momentum operators can be satisfied without necessarily realizing the canonical commutation relations.

Wigner's construction requires, in the simplest case (of a single bosonic oscillator), the Hamiltonian to be expressed as the anticommutator of the oscillators $a=a^-$ and $a^\dag=a^+$,
\begin{equation}
H= \frac{1}{2}\{a^-,a^+\},
\end{equation}
and, to compatibilize the Heisenberg equations of motion with the Hamilton equations, 
\begin{equation}
[ H,a^\pm ] = \pm a^\pm.
\end{equation}

Introducing additionally 
\begin{equation}
E^\pm=\{ a^\pm,a^\pm \} ,
\end{equation}
we obtain the orthosymplectic superalgebra $osp(1|2)$ (see \cite{fss}).  

Note that in this construction the creation and annihilation operators have odd nature, opposed to their bosonic physical nature\footnote{The same happens in BRST supersymmetry.}, and it will be precisely this fact that will solve the problem of (\ref{cope}). 

To solve the problem of the incompatibility of the deformation of the second-quantized fields with the deformation of the Heisenberg algebra, we start by rewriting the position, momentum and number operators in a Weyl-ordered form \cite{weyl}:
\begin{eqnarray}
  \tilde{X}_i &=& \frac{1}{2} \int d^Dy\; y_i \left( \psi^\dag(\vec{y})\psi(\vec{y}) + \psi(\vec{y})\psi^\dag(\vec{y})  \right) \label{n1} \\
  \tilde{P}_i &=& \frac{1}{2} \int d^Dp\; p_i \left( a^\dag_{\vec{p}} a_{\vec{p}} + a_{\vec{p}} a^{\dag}_{\vec{p}} \right)\\
  \tilde{N} &=& \frac{1}{2} \int d^Dy\;  \left( \psi^\dag(\vec{y})\psi(\vec{y}) + \psi(\vec{y})\psi^\dag(\vec{y})  \right).\label{n3}
\end{eqnarray}

They satisfy the same algebra as the operators  $X_i$, $P_i$ e $N$ previously defined. We have also rewritten the operator  $P_i$ in momentum space because it takes a diagonal form which will be of future convenience. 

As anticipated, we now declare that $\psi(\vec{y})$ and $a_{\vec{p}}$ are odd, and the coproduct of  $\tilde{X}_i$ is correctly induced as
\begin{eqnarray}
  \nonumber \Delta(\tilde{X}_i)&=&\frac{1}{2}\int d^Dy\; y_i (\Delta(\psi^\dag(\vec{y})) \Delta(\psi(\vec{y}))+\Delta(\psi(\vec{y})) \Delta(\psi^\dag(\vec{y})))=\\
  \nonumber &=&\frac{1}{2}\int d^Dy\; y_i [\psi^\dag(\vec{y})\psi(\vec{y})\otimes\mathbf{1}-
  \psi(\vec{y})\otimes\psi^\dag(\vec{y})+
  \psi^\dag(\vec{y})\otimes\psi(\vec{y})+\mathbf{1}\otimes\psi^\dag(\vec{y})\psi(\vec{y})+\\  \nonumber &&+\psi(\vec{y})\psi^\dag(\vec{y})\otimes\mathbf{1}-
  \psi^\dag(\vec{y})\otimes\psi(\vec{y})+\psi(\vec{y})\otimes\psi^\dag(\vec{y})+
  \mathbf{1}\otimes\psi(\vec{y})\psi^\dag(\vec{y})] =\\
  &=& \tilde{X}_i\otimes\mathbf{1}+\mathbf{1}\otimes \tilde{X}_i, 
\end{eqnarray}
the same holding, by a similar calculation, for the coproducts of  $\tilde{P}_i$ and $\tilde{N}$.

The antipode of $X_i$ is, as expected, 
\begin{eqnarray}
  \nonumber S(\tilde{X}_i)&=& \frac{1}{2}\int d^Dy\; y_i (S(\psi(\vec{y})\psi^\dag(\vec{y}))S(\psi^\dag(\vec{y})\psi(\vec{y}))=\\
  \nonumber &=& \frac{1}{2}\int d^Dy\; y_i[(-1)^{|\psi(\vec{y})||\psi^\dag(\vec{y})|}S(\psi^\dag(\vec{y}))
  S(\psi(\vec{y}))+(-1)^{|\psi^\dag(\vec{y})||\psi(\vec{y})|}S(\psi(\vec{y}))
  S(\psi^\dag(\vec{y}))] =\\
  &=&\frac{1}{2} \int d^Dy\; y_i \left( -\psi^\dag(\vec{y})\psi(\vec{y}) - \psi(\vec{y})\psi^\dag(\vec{y})\right)=
  \nonumber \\
  &=&-\tilde{X}_i,
\end{eqnarray}
as well as
\begin{eqnarray}
    S(\tilde{P}_i)=-\tilde{P}_i\\
    S(\tilde{N})=-\tilde{N}.
\end{eqnarray}

Counit poses no problem, since
$\epsilon(\psi(\vec{y}))=\epsilon(a_{\vec{p}})=0$ leads directly to  $\epsilon(\tilde{X}_i)=\epsilon(\tilde{P}_i)=\epsilon(\tilde{N})=0$.

The undeformed case is therefore solved. It remains to be shown that the deformations   (\ref{a1})-(\ref{a2}) and (\ref{de1})-(\ref{de2}) induce correctly the deformations of the bilinear integrated objects. 

The deformation of $\tilde{X}_i$ can be easily calculated in momentum space:
\begin{equation}\label{xtil}
      \tilde{X}^\mathcal{F}_i = \frac{i\hbar}{4} \int d^Dp \;\left(a^{\dag\mathcal{F}}_{\vec{p}}\stackrel{\leftrightarrow}{\partial_{p_i}}a^\mathcal{F}_{\vec{p}}+a^\mathcal{F}_{\vec{p}}\stackrel{\leftrightarrow}{\partial_{p_i}}a^{\dag\mathcal{F}}_{\vec{p}}\right)=\tilde{X}_i-\frac{1}{2}\alpha_{ij}p_j\hbar\tilde{N},
\end{equation}
and it is easy to see that it reduces to  (\ref{def}) at the one-particle limit ($\hbar \tilde{N}\rightarrow \hat{\hbar}$).

The (absence of) deformation of $\tilde{P}_i$ can also be easily calculated
\begin{eqnarray}
  \tilde{P}_i^\mathcal{F} &=& \frac{1}{2} \int d^Dp\; p_i \left( a^{\dag\mathcal{F}}_{\vec{p}} a^\mathcal{F}_{\vec{p}} + a^\mathcal{F}_{\vec{p}} a^{\dag\mathcal{F}}_{\vec{p}} \right)=\nonumber \\
  &=& \frac{1}{2} \int d^Dp\; p_i \left( a^\dag_{\vec{p}}\,e^{\frac{i}{2}\alpha_{ij}p_iP_j}e^{\frac{-i}{2}\alpha_{ij}p_iP_j} a_{\vec{p}} + a_{\vec{p}}\,e^{\frac{-i}{2}\alpha_{ij}p_iP_j}e^{\frac{i}{2}\alpha_{ij}p_iP_j} a^{\dag}_{\vec{p}} \right)  =\nonumber\\
  &=& \tilde{P}_i.
\end{eqnarray}

We can thus conclude that it is possible to construct the universal enveloping algebra of the algebra of the Schr\"odinger fields and oscillators and deform it by means of a Drinfel'd twist. Moreover, it is also possible to correctly induce the Hopf algebra structure of the position and momentum operators, both in the undeformed and in the deformed case, as long as the basic fields and oscillators are taken to be odd generators of a Lie superalgebra.

\section{Extended Heisenberg Algebras}

In this section we present a very simple example of a procedure that can have interesting applications. The procedure consists of constructing composite elements from the primitive elements of the Heisenberg algebra and then, for physical reasons (e.g. the necessity of correctly inducing a compostion rule for multiparticle states), declare that they are primitive elements of an enlarged algebra. The original composite nature is used only to compute the structure constants of the new algebra. These ideas are discussed in detail in \cite{ckt}.

We start with the Heisenberg algebra $\mathfrak{h}(\mathcal{N})$ and introduce the elements
\begin{eqnarray}
  K_{ij} &=& \frac{p_ip_j}{\hbar}, \nonumber\\
  M_{ij} &=& \frac{x_ip_j}{\hbar}, \nonumber\\
    N_{ij} &=& \frac{p_ix_j}{\hbar}, \nonumber\\
      V_{ij} &=& \frac{x_ix_j}{\hbar},
\end{eqnarray}
which we now declare to be primitive elements of an enlarged algebra. 

The enlarged algebra satisfies
\begin{eqnarray}
  \left[K_{ij}, x_k\right] &=& -i\delta_{ik}p_j-i\delta_{jk}p_i ,\nonumber \\
  \left[M_{ij}, x_k\right] &=& -i\delta_{jk}x_i ,\nonumber\\
  \left[N_{ij},x_k\right] &=& -i\delta_{ik}x_j ,\nonumber\\
  \left[M_{ij},p_k\right] &=& i\delta_{ik}p_j ,\nonumber\\
  \left[N_{ij},p_k\right] &=& i\delta_{jk}p_i ,\nonumber\\
  \left[V_{ij},p_k\right] &=& i\delta_{ik}x_j+i\delta_{jk}x_i ,\nonumber\\
  \left[V_{ij},K_{kl}\right] &=& i\delta_{jk}M_{il}+i\delta_{jl}M_{ik}+i\delta_{ik}N_{lj}+i\delta_{il}N_{kj} ,\nonumber\\
  \left[V_{ij},M_{kl}\right] &=& i\delta_{il}V_{jk}+i\delta_{jl}V_{ik} ,\nonumber\\
  \left[V_{ij},N_{kl}\right] &=& i\delta_{ik}V_{jl}+i\delta_{jk}V_{il} ,\nonumber\\
  \left[K_{ij},M_{kl}\right] &=& -i\delta_{ik}K_{jl}-i\delta_{jk}K_{il} ,\nonumber\\
  \left[K_{ij},N_{kl}\right] &=& -i\delta_{il}K_{jk}-i\delta_{jl}K_{ik}, \nonumber\\
  \left[M_{ij},N_{kl}\right] &=& i\delta_{ik}M_{lj}-i\delta_{jl}M_{ik}.
\end{eqnarray}
It is, in fact, the semidirect sum $\mathfrak{h}(\mathcal{N})\oplus_s sp(2\mathcal{N})$ \cite{val}.

We can now consider the Hamiltonian of the harmonic oscillator given by
\begin{equation}
    H=\sum_i \frac{p_i^2}{2}+\omega^2\sum_i\frac{x_i^2}{2}=\lambda\left(K_{ii}+\omega^2V_{ii} \right),
\end{equation}
where $\lambda$ is a suitable normalization constant.

We now apply the usual twist
\begin{equation}
    \mathcal{F}=\exp(i\alpha_{ij}p_i\otimes p_j),
\end{equation}
with $\alpha_{ij}=-\alpha_{ji}$.

The deformed Hamiltonian is
\begin{equation}
    H^\mathcal{F}=H-2\lambda\omega^2\hbar\alpha_{ij}M_{ij}+\lambda\omega^2\hbar^2\alpha_{ij}\alpha_{ik}K_{jk}.
\end{equation}

The deformed coproduct of the Hamiltonian is
\begin{equation}
    \Delta^{\cal F}(H)=\Delta(H)-2\lambda\omega^2\alpha_{ij}(p_i\otimes x_j-x_j\otimes p_i)+\lambda\omega^2\alpha_{ij}\alpha_{kj}(\hbar K_{ik}\otimes\hbar-\hbar\otimes \hbar K_{ik}),
\end{equation}
while the deformed coproduct of the deformed Hamiltonian is
\begin{equation}
\Delta^\mathcal{F}(H^\mathcal{F})=H^\mathcal{F}\otimes \mathbf{1}+\mathbf{1}\otimes
H^\mathcal{F}-4\lambda\omega^2\alpha_{ij}(p_i\otimes x_j)+2\lambda\omega^2\alpha_{ij}\alpha_{kj}(\hbar K_{ik}\otimes\hbar).
\end{equation}

This kind of extension of the Heisenberg algebra and the class of Hamiltonians that it yields have been discussed, for instance, in the context of conformal mechanics and bosonized supersymmetry \cite{cop}.

\chapter{Fermionic Heisenberg Algebras and Twisted Supersymmetric Quantum Mechanics}

Quantum theories in nonanticommutative spaces have been studied in the Drinfel'd twist approach (\cite{zup}, \cite{drw}). 
In this chapter we will study the twist deformations of the one-dimensional ${\cal N}$-extended supersymmetric
quantum mechanics. We will show that two constructions are possible: the identification of the supersymmetry algebra with the fermionic Heisenberg algebra (possible for even ${\cal N}$), and the realization of the supersymmetry algebra in terms of operators belonging to the universal enveloping algebra generated by one bosonic and ${\cal N}$ fermionic oscillators (possible for every ${\cal N}$). We shall recover, in a more general context, some results of the literature.

\section{Fermionic Heisenberg Algebra and One-dimensional ${\cal N}$-extended Supersymmetry}

Consider the Grassmann algebra generated by $\mathcal{N}$ anticommuting coordinates $\theta_\alpha$.  These coordinates, along with their Berezin derivatives $\partial_\alpha=\frac{\partial}{\partial \theta_\alpha}$, form a Lie superalgebra with $2\mathcal{N}$ odd generators and one single even generator, the central extension $z$. It is the fermionic Heisenberg algebra  $\mathfrak{h}_F(\mathcal{N})$ satisfying the (anti)commutatation relations 
\begin{eqnarray}
  \left\{\theta_\alpha,\theta_\beta\right\}&=&\left\{\partial_\alpha,\partial_\beta\right\}=0, \label{gras1}\\
  \left\{\partial_\alpha,\theta_\beta\right\}&=&\delta_{\alpha\beta}z,\\
  \left[z,\partial_\alpha\right]&=&\left[z,\theta_\alpha\right]=0. \label{gras3}
\end{eqnarray}

As discussed before, a careful interpretation of the role of the central extension is necessary, its coproduct and antipode being given by
\begin{eqnarray}
 \Delta(z)&=&z\otimes\mathbf{1}+\mathbf{1}\otimes z\\
 S(z)&=&-z.
\end{eqnarray}

Mass dimensions $[\theta_\alpha]=-\frac{1}{2}$, $[\partial_\alpha]=\frac{1}{2}$, $[z]=0$ can be attributed to the generators of $\mathfrak{h}_F(\mathcal{N})$.

We now go to the universal enveloping algebra $\mathcal{U}(\mathfrak{h}_F(\mathcal{N}))$, which has a Hopf algebra structure, and deform it by means of the Abelian twist 
\begin{equation}\label{abtwist}
    \mathcal{F}= \exp\left(C_{\alpha\beta}\partial_\alpha\otimes\partial_\beta \right),
    \end{equation}
expressed in terms of the diagonal matrix
\begin{equation}
C_{\alpha\beta}=\frac{1}{M}\eta_{\alpha\beta},
\end{equation}
where $M$ is a mass parameter and $\eta_{\alpha\beta}$ is a dimensionless diagonal matrix admitting, without loss of generality, $p$ entries $+1$ , $q$  entries  $-1$ and $r$ zero entries ($p+q+r=\mathcal{N}$).

With the techniques developed in the previous chapters, we obtain the deformation of the generators
$\theta_\alpha$ 
\begin{equation} \label{defteta}
    \theta_\alpha^\mathcal{F}={\overline f}^\beta(\theta_\alpha){\overline f}_\beta=\theta_\alpha+C_{\alpha\beta}\partial_\beta z,
\end{equation}
while the others undergo no deformation because they (anti)commute with the  $\partial_\alpha$s of the twist. The deformed generators differ from the original ones by the fermionic counterpart of the Bopp shift \cite{bopp}.

The deformed coproduct of $\theta_\alpha$  is
\begin{eqnarray}
    \Delta^\mathcal{F}(\theta_\alpha)&=&\exp\left(C_{\alpha\beta}\partial_\alpha\otimes\partial_\beta \right)\Delta(\theta_\alpha)\exp\left(-C_{\alpha\beta}\partial_\alpha\otimes\partial_\beta \right)\nonumber\\
    &=&\theta_\alpha\otimes\mathbf{1} + \mathbf{1}\otimes\theta_\alpha+C_{\alpha\beta}(\partial_\beta\otimes z -
    z\otimes\partial_\beta),
\end{eqnarray}
and, obviously,
\begin{eqnarray}
    \Delta^\mathcal{F}(\partial_\alpha)&=&\partial_\alpha\otimes\mathbf{1} + \mathbf{1}\otimes\partial_\alpha\\
    \Delta^\mathcal{F}(z)&=&z\otimes\mathbf{1} + \mathbf{1}\otimes z.
\end{eqnarray}

Since
\begin{equation}
    \chi=f^\alpha
    S(f_\alpha)=\exp\left(-C_{\alpha\beta}\partial_\alpha\partial_\beta\right)=\mathbf{1},
\end{equation}
the antipode does not get deformed.

The universal R-matrix is given by  
\begin{equation}
    \mathcal{R}=\sum_{\alpha,\beta}(-1)^{|\bar{f}^\beta||f^\alpha|}(f_\alpha\bar{f}^\beta\otimes f^\alpha\bar{f}_\beta)=\exp\left(-2C_{\alpha\beta}\partial_\alpha\otimes\partial_\beta \right),
\end{equation}
so that we obtain the deformed coproduct of $\theta_\alpha^\mathcal{F}$ as  
\begin{equation}
    \Delta^\mathcal{F}(\theta_\alpha^\mathcal{F})=
    \theta_\alpha^\mathcal{F}\otimes\mathbf{1}+\mathbf{1}\otimes\theta_\alpha^\mathcal{F}+2C_{\alpha\beta}\partial_\beta\otimes
    z.
\end{equation}

With the antipode 
\begin{equation}
    S(\theta_\alpha^\mathcal{F})=-\theta_\alpha+C_{\alpha\beta}z\partial_\beta
    =-\theta_\alpha^\mathcal{F}+2C_{\alpha\beta}z\partial_\beta
\end{equation}
in hand we can calculate the deformed brackets
\begin{eqnarray}
  \left\{\theta_\alpha^\mathcal{F},\partial_\beta^\mathcal{F}\right\}_\mathcal{F}
  &=&\delta_{\alpha\beta}z^\mathcal{F}, \nonumber\\
  \left\{\theta_\alpha^\mathcal{F},\theta_\beta^\mathcal{F}\right\}_\mathcal{F}&=&0, \\
  \left\{\partial_\alpha^\mathcal{F},\partial_\beta^\mathcal{F}\right\}_\mathcal{F}&=&0, \\
  \left[\partial_\alpha^\mathcal{F},z^\mathcal{F}\right]_\mathcal{F}
  &=&\left[\theta_\alpha^\mathcal{F},z^\mathcal{F}\right]_\mathcal{F}=0,
\end{eqnarray}
which have the same structure constants as the original algebra (\ref{gras1})--(\ref{gras3}).

The ordinary brackets of the deformed generators make the formerly Grassmannian generators nonanticommuting:
\begin{eqnarray}
  \left\{\theta_\alpha^\mathcal{F},\partial_\beta^\mathcal{F}\right\}&=&\delta_{\alpha\beta}z, \\
  \left\{\theta_\alpha^\mathcal{F},\theta_\beta^\mathcal{F}\right\}&=&2C_{\alpha\beta}z^2, \\
  \left\{\partial_\alpha^\mathcal{F},\partial_\beta^\mathcal{F}\right\}&=&0, \\
  \left[z^\mathcal{F},\theta_\alpha^\mathcal{F}\right]
  &=&\left[z^\mathcal{F},\partial_\alpha^\mathcal{F}\right]=0.  \label{d4}
\end{eqnarray}

We shall now study the deformation of the multiplication $m$ on a module $M$. The module will be again a space of functions, but this time of the Grassmann variables, and the action of  $\mathfrak{h}_F(\mathcal{N})$ will be given by 
\begin{eqnarray}
    \partial_\alpha\triangleright a = z\frac{\partial a}{\partial\theta_\alpha} \\
    \theta_\alpha\triangleright a = \theta_\alpha\cdot a,
\end{eqnarray}
where $\cdot$ denotes usual Grassmannian multiplication, i.e., $a\cdot b=m(a\otimes b)$, $a,b\in M$.

Since we are working on a module, which furnishes a representation of $\mathfrak{h}_F(\mathcal{N})$, a numerical value can be attributed to the central extension. For convenience, we are setting $z=1$. 

The deformed multiplication is
\begin{equation}
    a\star b\equiv m^\mathcal{F}(a\otimes b)=(m\circ\mathcal{F}^{-1})(a\otimes b).
\end{equation}

Defining $\left[a,b\right\}_\star\equiv a\star b
+(-1)^{|a||b|} b\star a$, we have 
\begin{eqnarray}
  \left\{ \theta_\alpha , \theta_\beta\right\}_\star&=&2C_{\alpha\beta}, \\
  \left\{\partial_\alpha , \theta_\beta\right\}_\star&=&\delta_{\alpha\beta}, \\
  \left\{\partial_\alpha , \partial_\beta\right\}_\star&=&0,
\end{eqnarray}
which is a cliffordization of the Grassmann coordinates, similar to the obtained in \cite{hh}, \cite{hh2}, \cite{hh3} .

We now consider the algebra of the one-dimensional $\mathcal{N}$-extended supersymmetry and show that, for even values of  $\mathcal{N}$, it is isomorphic to $\mathfrak{h}_F(\frac{\mathcal{N}}{2})$, so that the deformation obtained above can be easily applied if we make the appropriate identifications. 

Consider the algebra of the supersymmetry generators  ${\widehat Q}_I$  and the central extension $H$ satisfying
\begin{eqnarray}\label{susy}
  \{\widehat{Q}_I, \widehat{Q}_J\} &=& \delta_{IJ}H,\\
    \left[H,\widehat{Q}_I\right] &=& 0,
\end{eqnarray}
$I,J=1,\ldots, {\cal N}$.

For ${\cal N}$ even, we can split the odd generators into the chiral sector  $Q_i$ and antichiral sector $\overline{Q}_i$:
\begin{eqnarray}
  Q_i &=& \widehat{Q}_i+i\widehat{Q}_{i+\frac{{\cal N}}{2}},\\
  \overline{Q}_i &=& \widehat{Q}_i-i\widehat{Q}_{i+\frac{{\cal N}}{2}},
  \end{eqnarray}
with $i=1,\ldots,\frac{{\cal N}}{2}$.

The algebra can then be reexpressed as
\begin{eqnarray}\label{susy2}
\left\{Q_i,\overline{Q}_j\right\}&=&2\delta_{ij}H,\\
  \left\{Q_i,Q_j\right\}&=&\left\{\overline{Q}_i,\overline{Q}_j\right\}=0,\\
  \left[H,Q_i\right]&=&\left[H,\overline{Q}_i\right]=0.\label{susy3}
\end{eqnarray}

This algebra is isomorphic to (\ref{gras1})-(\ref{gras3}) if we identify

\begin{eqnarray}
Q_i &\leftrightarrow&\theta_\alpha \\
 \overline{Q}_i &\leftrightarrow& \partial_\alpha\\
2H  &\leftrightarrow& z.
\end{eqnarray}

Using this identification in (\ref{abtwist}), we will now deform the algebra (\ref{susy2})-(\ref{susy3}) by means of the Abelian twist 
\begin{equation}\label{339}
    \mathcal{F}=\exp\left(\frac{C_{ij}}{2}\overline{Q}_i\otimes\overline{Q}_j \right),
\end{equation}
with $C_{ij}=\frac{\eta_{ij}}{M}$, where $\eta_{ij}$ is a dimensionless diagonal matrix with $p$ positive entries, $q$ negative entries and $r$ zero entries ($p+q+r={\cal N}$) and $M$ is a mass parameter.

This deformation, as anticipated, coincides with (\ref{abtwist}). 

The deformed coproduct of ${Q_i}$ is
\begin{equation}
    \Delta^\mathcal{F}({Q}_i)=\Delta({Q}_i)+C_{ij}(\overline{Q}_j\otimes H-H\otimes \overline{Q}_j).
\end{equation}

The only deformed generators are the $Q_i$s, whose deformation is given by
\begin{equation}
    {Q}_i^\mathcal{F}=Q_i+C_{ij}\overline{Q}_j
    H.
\end{equation}

The universal R-matrix  is $\mathcal{F}^{-2}$, so that the deformed coproduct of the deformed generators is
\begin{equation}
    \Delta^\mathcal{F}(Q_i^\mathcal{F})=Q_i^\mathcal{F}\otimes\mathbf{1}+\mathbf{1}\otimes Q_i^\mathcal{F}+2C_{ij}{\overline{Q}}_j\otimes
    H,
\end{equation}
which, along with the antipode 
\begin{equation}
    S(Q_i^\mathcal{F})=-Q_i+C_{ij}\overline{Q}_j
    H=-Q_i^\mathcal{F}+2C_{ij}\overline{Q}_j
    H,
\end{equation}
allows the calculation of the deformed brackets
\begin{eqnarray}
  \left\{\overline{Q}_i^\mathcal{F},Q_j^\mathcal{F}\right\}_\mathcal{F}&=&\delta_{ij}H^\mathcal{F}=\delta_{ij}H,\\
  \left\{\overline{Q}_i^\mathcal{F},\overline{Q}_j^\mathcal{F}\right\}_\mathcal{F}&=&0,\\
  \left\{Q_i^\mathcal{F},Q_j^\mathcal{F}\right\}_\mathcal{F}&=&0,\\
  \left[Q_i^\mathcal{F},H^\mathcal{F}\right]_\mathcal{F}&=&\left[\overline{Q}_i^\mathcal{F},H^\mathcal{F}\right]_\mathcal{F}=0.
\end{eqnarray}

The ordinary brackets of the deformed quantities are
\begin{eqnarray}\label{mixsusy}
  \left\{\overline{Q}_i^\mathcal{F},Q_j^\mathcal{F}\right\}&=&\delta_{ij}H^\mathcal{F},\\
  \left\{\overline{Q}_i^\mathcal{F},\overline{Q}_j^\mathcal{F}\right\}&=&0, \\
  \left\{Q_i^\mathcal{F},Q_j^\mathcal{F}\right\}&=&2C_{ij}({H^{\mathcal{F}}})^2, \\
  \left[H^\mathcal{F},\overline{Q}_i^\mathcal{F}\right]&=&\left[H^\mathcal{F},Q_i^\mathcal{F}\right]=0.
\end{eqnarray}

Note that nonlinear superalgebras of the form $\{Q_a, Q_b\}=\delta_{ab}P_n(H)$, where $P_n(H)$ is a degree $n$ polynomial of the  Hamiltonian, were introduced in  \cite{ais}.

\section{Superspace Representation}

We now want to study the twist deformation of supersymmetric quantum mechanics in the superspace representation. 
It is necessary to introduce the set of Grassmann variables $\theta_I$ and their derivatives $\partial_{\theta_I}$,
which, along with the central extension $z$, form the algebra $\mathfrak{h}_F(N)$, as well as the bosonic parameter $t$ and its derivative $\partial_t$, which, along with the central element ${\hbar}$, satisfy the bosonic Heisenberg algebra $\mathfrak{h}(1)$. We obtain, in principle, the algebra $\mathfrak{h}(1)\oplus \mathfrak{h}_F(N)$. We can now identify the central extensions ($z=\hbar$), thus obtaining an algebra which we shall call  $\mathfrak{h}(1,N)$. The algebra of the $\mathcal{N}$-extended supersymmetry can be explicitly realized in terms of operators of the algebra ${\cal U}(\mathfrak{h}(1,\mathcal{N}))$:
\begin{eqnarray}\label{q}
    \widehat{Q}_I &=& \partial_{\theta_I}+\frac{i} {\hbar}{\theta_I\partial_t},\\
    H &=& i\partial_t,
\end{eqnarray}
with $I=1,\ldots,{\cal N}$.

Here it is necessary to impose, along the lines of what was discussed in section \textbf{2.3} and in  \cite{ckt}, that the undeformed coproduct of $\widehat{Q}_I$ coincides with the coproduct of a primitive fermionic generator, i.e., $\Delta({\widehat Q}_I)=\widehat{Q}_I\otimes \mathbf{1}+\mathbf{1}\otimes\widehat{Q}_I$, so that the presence of terms of the form $\frac{1}{\hbar}$ in the expression above is not problematic.

Since we are working in ${\cal U}(\mathfrak{h}(1,\mathcal{N}))$, it is now possible to deform the supersymmetry generators by means of the twist (\ref{abtwist})
\begin{equation}
\mathcal{F}=\exp(C_{IJ}\partial_{\theta_I}\otimes\partial_{\theta_J}),
\end{equation}
obtaining as deformed generators
\begin{equation}
    \widehat{Q}_I^\mathcal{F}=\widehat{Q}_I+C_{IJ}H\partial_{\theta_J}
\end{equation}
and  $H^{\cal F}=H$.

The deformed coproduct of   $\widehat{Q}_I$ is
\begin{equation}
    \Delta^\mathcal{F}(\widehat{Q}_I)=\Delta(\widehat{Q}_I)+C_{IJ} (\partial_{\theta_J}\otimes H-H\otimes\partial_{\theta_J}),
    \end{equation}
while the deformed coproduct of $\widehat{Q}_I^{\cal F}$ is
\begin{equation}
    \Delta^\mathcal{F}(\widehat{Q}_I^\mathcal{F})=\widehat{Q}_I^\mathcal{F}\otimes \mathbf{1}+\mathbf{1}\otimes\widehat{Q}_I^\mathcal{F}+2C_{IJ} (\partial_{\theta_J}\otimes H).
    \end{equation}

The ordinary brackets of the deformed generators are
        \begin{equation}\label{nlinsusy2}
        \{\widehat{Q}_I^\mathcal{F},\widehat{Q}_J^\mathcal{F}\}=\delta_{IJ}H+2C_{IJ}H^2,
    \end{equation}
while the deformed brackets, as expected, restore the original algebra
   \begin{equation}
        \{\widehat{Q}_I^\mathcal{F},\widehat{Q}_J^\mathcal{F}\}_\mathcal{F}=\delta_{IJ}H.
    \end{equation}
    
We will now discuss the particular cases of ${\cal N}=2$ and ${\mathcal{N}=4}$.

\subsection{The ${\cal N}=2$ case}

We consider here the realization of the ${\cal N}=2$ supersymmetry generators in terms of elements of the algebra ${\cal U}(\mathfrak{h}(1,2))$:
\begin{eqnarray}
  \widehat{Q}_1 &=& \partial_{\theta_1}+\frac{i}{\hbar}\theta_1\partial_t, \\
  \widehat{Q}_2 &=& \partial_{\theta_2}+\frac{i}{\hbar}\theta_2\partial_t.
  \end{eqnarray}
The fermionic derivatives
  \begin{eqnarray}
  D_1 &=& \partial_{\theta_1}-\frac{i}{\hbar}\theta_1\partial_t,  \\
  D_2 &=& \partial_{\theta_2}-\frac{i}{\hbar}\theta_2\partial_t
\end{eqnarray}
can be added.

These four generators, together with the central extension $H$, satisfy the so-called ${\cal N}=(2,2)$ pseudosupersymmetry algebra: 
\begin{eqnarray}
  \{\widehat{Q}_I, \widehat{Q}_J\} &=& \delta_{IJ}H,\\
  \{D_I,D_J\} &=& -\delta_{IJ}H,\\
  \{D_I,\widehat{Q}_J\} &=& 0, \\
  \left[H, \widehat{Q}_I\right] &=& \left[H, D_I\right]=0.
\end{eqnarray}

We can now deform this algebra with any twist $\mathcal{F}\in\mathcal{U}(\mathfrak{h}(1,2))\otimes\mathcal{U}(\mathfrak{h}(1,2))$ which is invertible and satisfies the 2-cocycle condition. In particular, an admissible Abelian twist is 
\begin{equation}
    \mathcal{F}=\exp\left(\frac{\epsilon}{M}\overline{Q}\otimes\overline{Q}+\frac{\eta}{M}\overline{D}\otimes\overline{D} \right),
\end{equation}
where
\begin{eqnarray}
  \overline{Q} &=& \widehat{Q}_1-i\widehat{Q}_2, \\
  \overline{D} &=& D_1-iD_2
\end{eqnarray}
and $\epsilon$, $\eta$ are numbers normalizable to  $+1$, $-1$ or $ 0$ without loss of generality. The twist (\ref{339}) is recovered for $\eta=0$, $C_{11}=\frac{\epsilon}{M}$. 

The deformed generators are
\begin{eqnarray}
  \widehat{Q}_1^\mathcal{F} &=& \widehat{Q}_1+\frac{\epsilon}{M}(\widehat{Q}_1-i\widehat{Q}_2)H, \\
  \widehat{Q}_2^\mathcal{F} &=& \widehat{Q}_2-\frac{i\epsilon}{M}(\widehat{Q}_1-i\widehat{Q}_2)H, \\
  D_1^\mathcal{F} &=& D_1+\frac{\eta}{M}(D_1-iD_2)H, \\
  D_2^\mathcal{F} &=& D_2-\frac{i\eta}{M}(D_1-iD_2)H,
\end{eqnarray}
and $H^{\mathcal F}=H$.

The deformed coproducts are given by
\begin{eqnarray}
  \Delta^\mathcal{F}(\widehat{Q}_1) &=& \Delta(\widehat{Q}_1)+\frac{\epsilon}{M}(\widehat{Q}_1\otimes H-H\otimes \widehat{Q}_1)-\frac{i\epsilon}{M}(\widehat{Q}_2\otimes H-H\otimes \widehat{Q}_2),\nonumber \\
   \Delta^\mathcal{F}(\widehat{Q}_2) &=& \Delta(\widehat{Q}_2)-\frac{i\epsilon}{M}(\widehat{Q}_1\otimes H-H\otimes \widehat{Q}_1)-\frac{\epsilon}{M}(\widehat{Q}_2\otimes H-H\otimes \widehat{Q}_2),\nonumber\\
   \Delta^\mathcal{F}(D_1) &=& \Delta(D_1)-\frac{\eta}{M}(D_1\otimes H-H\otimes D_1)+\frac{i\eta}{M}(D_2\otimes H-H\otimes D_2),\nonumber\\
 \Delta^\mathcal{F}(D_2) &=& \Delta(D_2)+\frac{i\eta}{M}(D_1\otimes H-H\otimes D_1)+\frac{\eta}{M}(D_2\otimes H-H\otimes D_2).\nonumber \\
 \;
\end{eqnarray}

Since
\begin{equation}
    \chi=f^\alpha S(f_\alpha)=\exp\left(-\frac{\epsilon}{M} \overline{Q}^2-\frac{\eta}{M} \overline{D}^2 \right)=\mathbf{1},
\end{equation}
the antipodes remain undeformed and are
\begin{eqnarray}
  S(\widehat{Q}_1^\mathcal{F}) &=& -\widehat{Q}_1^\mathcal{F}+\frac{2\epsilon}{M}(\widehat{Q}_1-i\widehat{Q}_2),  \\
  S(\widehat{Q}_2^\mathcal{F}) &=& -\widehat{Q}_2^\mathcal{F}-\frac{2i\epsilon}{M}(\widehat{Q}_1-i\widehat{Q}_2),  \\
  S(D_1^\mathcal{F}) &=& -D_1^\mathcal{F}-\frac{2\eta}{M}(D_1-iD_2),   \\
  S(D_2^\mathcal{F}) &=& -D_2^\mathcal{F}+\frac{2i\eta}{M}(D_1-iD_2).
\end{eqnarray}

The universal R-matrix is simply
\begin{equation} \mathcal{R}=\mathcal{F}^{-2}=\exp\left[-2\left(\frac{\epsilon}{M}\overline{Q}\otimes\overline{Q}+\frac{\eta}{M}\overline{D}\otimes\overline{D} \right) \right].
\end{equation}

With these results, we can calculate the deformed coproducts of the deformed quantities. They are
\begin{eqnarray}
  \Delta^\mathcal{F}(\widehat{Q}_1^\mathcal{F}) &=& \widehat{Q}_1^\mathcal{F}\otimes\mathbf{1}+\mathbf{1}\otimes \widehat{Q}_1^\mathcal{F}+\frac{2\epsilon}{M}(\widehat{Q}_1-i\widehat{Q}_2)\otimes H, \\
  \Delta^\mathcal{F}(\widehat{Q}_2^\mathcal{F}) &=& \widehat{Q}_2^\mathcal{F}\otimes\mathbf{1}+\mathbf{1}\otimes \widehat{Q}_2^\mathcal{F}-\frac{2i\epsilon}{M}(\widehat{Q}_1-i\widehat{Q}_2)\otimes H,   \\
  \Delta^\mathcal{F}(D_1^\mathcal{F}) &=& D_1^\mathcal{F}\otimes\mathbf{1}+\mathbf{1}\otimes D_1^\mathcal{F}-\frac{2\eta}{M}(D_1-iD_2)\otimes H ,  \\
  \Delta^\mathcal{F}(D_2^\mathcal{F}) &=& D_2^\mathcal{F}\otimes\mathbf{1}+\mathbf{1}\otimes D_2^\mathcal{F}+\frac{2i\eta}{M}(D_1-iD_2)\otimes H,
\end{eqnarray}
and yield, as expected, the deformed brackets of the deformed generators with the original structure constants:
\begin{eqnarray}
  \{\widehat{Q}_I^\mathcal{F}, \widehat{Q}_J^\mathcal{F}\}_\mathcal{F} &=& \delta_{IJ}H^\mathcal{F},  \\
  \{D_I^\mathcal{F},D_J^\mathcal{F}\}_\mathcal{F} &=& -\delta_{IJ}H^\mathcal{F}, \\
  \{D_I^\mathcal{F},\widehat{Q}_J^\mathcal{F}\}_\mathcal{F} &=& 0.
\end{eqnarray}

We shall now study the deformed multiplication on a module which consists of the space of functions of the Grassmann variables
 $\theta_1, \theta_2$. The ordinary multiplication $m$ is the usual Grassmannian product, i.e.,
\begin{equation}
    m(\theta_I\otimes\theta_J)=\theta_I\cdot\theta_J.
\end{equation}

The action of $\widehat{Q}_I$ and $D_I$ on the module is given by
\begin{eqnarray}
  &\widehat{Q}_I\triangleright\theta_J=
  D_I\triangleright\theta_J=\delta_{IJ}.
\end{eqnarray}

We define the star product as 
\begin{equation}
    \theta_I\star\theta_J=m^\mathcal{F}(\theta_I\otimes\theta_J)=(m\circ\mathcal{F}^{-1})(\theta_I\otimes\theta_J)
\end{equation}
and calculate explicitly
\begin{eqnarray}
  \theta_1\star\theta_1&=& -\frac{\epsilon}{M} - \frac{\eta}{M}, \\
   \theta_2\star\theta_2&=& \frac{\epsilon}{M} + \frac{\eta}{M}, \\
    \theta_1\star\theta_2&=& -\frac{i\epsilon}{M} - \frac{i\eta}{M}+\theta_1\theta_2,
\end{eqnarray}
so that the star anticommutators are
\begin{eqnarray}
  \{\theta_1,\theta_1\}_\star&=& -2\left(\frac{\epsilon}{M} + \frac{\eta}{M}\right), \\
   \{\theta_2,\theta_2\}_\star&=& 2\left(\frac{\epsilon}{M} + \frac{\eta}{M}\right), \\
    \{\theta_1,\theta_2\}_\star&=& -2i \left(\frac{\epsilon}{M}+ \frac{\eta}{M}\right).
\end{eqnarray}

Going to chiral coordinates
\begin{eqnarray}
  \theta &=& \theta_1+i\theta_2,\nonumber \\
  \bar{\theta} &=& \theta_1-i\theta_2,
\end{eqnarray}
the star anticommutators are expressed as 
\begin{eqnarray}
  \{\theta,\theta\}_\star&=& -8\left(\frac{\epsilon}{M} + \frac{\eta}{M}\right), \\
   \{\bar{\theta},\bar{\theta}\}_\star&=& 0, \\
    \{\theta,\bar{\theta}\}_\star&=&0,
\end{eqnarray}
giving us the cliffordization in half the coordinates (chiral sector), as obtained in \cite{ks} and \cite{ihl}. 

Since $H^{\cal F}=H$, one could think that the bosonic sector does not suffer any deformation. However, consider the bosonic operator 
\begin{equation}
    W=\frac{i}{2}(\widehat{Q}_1\widehat{Q}_2-\widehat{Q}_2\widehat{Q}_1),
\end{equation}
which is also declared primitive, i.e., $\Delta(W)=W\otimes\mathbf{1}+\mathbf{1}\otimes W$ and $S(W)=-W$.

If we deform it (taking, for simplicity, $\eta=0$), we obtain that $W^\mathcal{F}=W$, because $[\overline{Q},W]=-2H\overline{Q}$. There is no deformation at the level of the algebra. The coproduct of $W$, however, is deformed:
\begin{equation}
    \Delta^\mathcal{F}(W)=\Delta(W)-\frac{2\epsilon}{M}(\overline{Q}\otimes\overline{Q}H+\overline{Q}H\otimes\overline{Q}),
\end{equation}
showing that the bosonic sector is not entirely immune to the fermionic twist.

\subsection{The ${\cal N}=4$ case}

We now turn our attention to the ${\cal N}=4$ supersymmetry algebra 
\begin{eqnarray}
  \{\widehat{Q}_I, \widehat{Q}_J\} &=&\delta_{IJ}H, \\
    \left[H,\widehat{Q}_I\right]&=&0,
\end{eqnarray}
($I,J=1,2,3,4$)
and apply the twist
\begin{equation}
    \mathcal{F}=\exp\left(\frac{\eta_{ij}}{M}\overline{Q}_i\otimes\overline{Q}_j \right),
\end{equation}
where $\eta_{ij}$ is diagonal and 
\begin{eqnarray}
  \overline{Q}_1 &=& \widehat{Q}_1-i\widehat{Q}_2, \\
  \overline{Q}_2 &=& \widehat{Q}_3-i\widehat{Q}_4.
\end{eqnarray}

We set $\eta_{11}=\epsilon$ e $\eta_{22}=\eta$. 

The same procedure allows us to calculate the deformed generators
\begin{eqnarray}
  \widehat{Q}_1^\mathcal{F} &=& \widehat{Q}_1+\frac{\epsilon}{M}(\widehat{Q}_1-i\widehat{Q}_2)H, \\
  \widehat{Q}_2^\mathcal{F} &=& \widehat{Q}_2-\frac{i\epsilon}{M}(\widehat{Q}_1-i\widehat{Q}_2)H,\\
  \widehat{Q}_3^\mathcal{F} &=& \widehat{Q}_3+\frac{\eta}{M}(\widehat{Q}_3-i\widehat{Q}_4)H, \\
  \widehat{Q}_4^\mathcal{F} &=& \widehat{Q}_4-\frac{i\eta}{M}(\widehat{Q}_3-i\widehat{Q}_4)H.
\end{eqnarray}

The deformed coproducts are
\begin{eqnarray}
  \Delta^\mathcal{F}(\widehat{Q}_1) &=& \Delta(\widehat{Q}_1)+\frac{\epsilon}{M}(\widehat{Q}_1\otimes H-H\otimes \widehat{Q}_1)-\frac{i\epsilon}{M}(\widehat{Q}_2\otimes H-H\otimes \widehat{Q}_2),\nonumber \\
   \Delta^\mathcal{F}(\widehat{Q}_2) &=& \Delta(\widehat{Q}_2)-\frac{i\epsilon}{M}(\widehat{Q}_1\otimes H-H\otimes \widehat{Q}_1)-\frac{\epsilon}{M}(\widehat{Q}_2\otimes H-H\otimes \widehat{Q}_2),\nonumber\\
   \Delta^\mathcal{F}(\widehat{Q}_3) &=& \Delta(\widehat{Q}_3)+\frac{\eta}{M}(\widehat{Q}_3\otimes H-H\otimes \widehat{Q}_3)-\frac{i\eta}{M}(\widehat{Q}_4\otimes H-H\otimes \widehat{Q}_4),\nonumber\\
 \Delta^\mathcal{F}(\widehat{Q}_4) &=& \Delta(\widehat{Q}_4)-\frac{i\eta}{M}(\widehat{Q}_3\otimes H-H\otimes \widehat{Q}_3)-\frac{\eta}{M}(\widehat{Q}_4\otimes H-H\otimes \widehat{Q}_4).\nonumber \\
 \;
\end{eqnarray}

The universal R-matrix is simply $\mathcal{F}^{-2}$, allowing us to calculate the deformed coproduct of the deformed generators \begin{eqnarray}
  \Delta^\mathcal{F}(\widehat{Q}_1^\mathcal{F}) &=& \widehat{Q}_1^\mathcal{F}\otimes\mathbf{1}+\mathbf{1}\otimes \widehat{Q}_1^\mathcal{F}+\frac{2\epsilon}{M}(\widehat{Q}_1-i\widehat{Q}_2)\otimes H,  \\
  \Delta^\mathcal{F}(\widehat{Q}_2^\mathcal{F}) &=& \widehat{Q}_2^\mathcal{F}\otimes\mathbf{1}+\mathbf{1}\otimes \widehat{Q}_2^\mathcal{F}-\frac{2i\epsilon}{M}(\widehat{Q}_1-i\widehat{Q}_2)\otimes H,    \\
  \Delta^\mathcal{F}(\widehat{Q}_3^\mathcal{F}) &=& \widehat{Q}_3^\mathcal{F}\otimes\mathbf{1}+\mathbf{1}\otimes \widehat{Q}_3^\mathcal{F}+\frac{2\eta}{M}(\widehat{Q}_3-i\widehat{Q}_4)\otimes H,    \\
  \Delta^\mathcal{F}(\widehat{Q}_4^\mathcal{F}) &=& \widehat{Q}_4^\mathcal{F}\otimes\mathbf{1}+\mathbf{1}\otimes \widehat{Q}_4^\mathcal{F}-\frac{2i\eta}{M}(\widehat{Q}_3-i\widehat{Q}_4)\otimes H .
\end{eqnarray}

The antipodes are
\begin{eqnarray}
  S(\widehat{Q}_1^\mathcal{F}) &=& -\widehat{Q}_1^\mathcal{F}+\frac{2\epsilon}{M}(\widehat{Q}_1-i\widehat{Q}_2),\\
  S(\widehat{Q}_2^\mathcal{F}) &=& -\widehat{Q}_2^\mathcal{F}-\frac{2i\epsilon}{M}(\widehat{Q}_1-i\widehat{Q}_2),  \\
  S(\widehat{Q}_3^\mathcal{F}) &=& -\widehat{Q}_3^\mathcal{F}+\frac{2\eta}{M}(\widehat{Q}_3-i\widehat{Q}_4),  \\
  S(\widehat{Q}_4^\mathcal{F}) &=& -\widehat{Q}_4^\mathcal{F}-\frac{2i\eta}{M}(\widehat{Q}_3-i\widehat{Q}_4),
\end{eqnarray}
so that the deformed brackets are, as expected,
\begin{eqnarray}
  \{\widehat{Q}_I^\mathcal{F}, \widehat{Q}_J^\mathcal{F}\}_\mathcal{F} &=&\delta_{IJ}H^\mathcal{F}, \\
    \left[H^\mathcal{F},\widehat{Q}_I^\mathcal{F}\right]_\mathcal{F}&=&0.
\end{eqnarray}

We now examine the deformed multiplication $m^\mathcal{F}$ on the space of functions of the Grassmann variables $\theta_I$, $I=1,\ldots,4$, with the action of the $\widehat{Q}_I$s on the module given by  $ \widehat{Q}_I\triangleright\theta_J= \delta_{IJ}$.

With the same definition of star product, we have
\begin{eqnarray}
 \theta_1\star\theta_1&=& -\frac{\epsilon}{M}, \\
   \theta_2\star\theta_2&=& \frac{\epsilon}{M}, \\
    \theta_3\star\theta_3&=& -\frac{\eta}{M}, \\
   \theta_4\star\theta_4&=& \frac{\eta}{M};
\end{eqnarray}
all the other combinations coincide with the usual Grassmannian product.

Going to chiral coordinates
\begin{eqnarray}
  \zeta_1 &=& \theta_1+i\theta_2, \\
  \overline{\zeta_1} &=& \theta_1-i\theta_2,\\
  \zeta_2 &=& \theta_3+i\theta_4,\\
  \overline{\zeta_2} &=& \theta_3-i\theta_4,
\end{eqnarray}
the star anticommutators are
\begin{eqnarray}
  \{\zeta_I,\zeta_J\}_\star&=& -8\frac{\eta_{IJ}}{M},\\
   \{\overline{\zeta}_I,\overline{\zeta}_J\}_\star&=& 0,\\
    \{\zeta_I,\overline{\zeta}_J\}_\star&=&0,
\end{eqnarray}
giving rise to the cliffordization of the unbarred chiral coordinates, as in \cite{ks} and \cite{ihl}.

To show that the bosonic sector is also affected by the twist, we proceed similarly and introduce the Hermitian bosonic operators 
 \begin{eqnarray}
    W_1 &=& \frac{i}{2}(\widehat{Q}_1\widehat{Q}_2-\widehat{Q}_2\widehat{Q}_1) ,\\
     W_2 &=& \frac{i}{2}(\widehat{Q}_3\widehat{Q}_4-\widehat{Q}_4\widehat{Q}_3),
\end{eqnarray}
which are again considered primitive elements.

Their algebra with the $\overline{Q}_i$s is
\begin{equation}
    [\overline{Q}_i,W_j]=-2H\overline{Q}_i\delta_{ij} \text{    (no sum over \emph{i})},
\end{equation}
so that there is no deformation at the algebric level, i.e, $W_i^\mathcal{F}=W_i$.  The same is not true at the coalgebric level, because their deformed coproducts are
\begin{equation}
\Delta^\mathcal{F}(W_i)=\Delta(W_i)-\frac{2\eta_{ij}}{M}(\overline{Q}_j\otimes\overline{Q}_jH+\overline{Q}_jH\otimes\overline{Q}_j).
\end{equation}

\addcontentsline{toc}{chapter}{Concluding Remarks}
\chapter*{Concluding Remarks}

In this work we proved that it is possible to construct the universal enveloping algebra $\mathcal{U}(\mathfrak{h})$ of the Heisenberg algebra and deform it by means of an Abelian Drinfel'd twist, as long as the role of the central extension is correctly taken into account, i.e., it is considered as a generic element of the Lie algebra and not a multiple of the identity. We showed that the deformed commutators of the deformed generators $x_i^{\mathcal F}$ and $p_i^{\mathcal F}$ of the linear subspace of $\mathcal{U}^{\mathcal F}(\mathfrak{h})$ exhibit the same structure constants of the original algebra, that is, the algebra $[x_i,x_j]=0$ is reproduced by the deformation: $[x_i^{\mathcal F},x_j^{\mathcal F}]_{\cal F}=0$. Noncommutativity emerges in the hybrid case, when we compute the ordinary commutators of the deformed generators as $[x_i^{\mathcal F},x_j^{\mathcal F}]=i\theta_{ij}$. We also recalled how to implement a star product on a module, and how this product also gives rise to noncommutativity of the form $[\check{x}_i,\check{x}_j]_\star=i\theta_{ij}$. 

Going to the second-quantization formalism, the generators of position and momentum of the Heisenberg algebra are realized as integrated bilinears of the Schr\"odinger fields and oscillators (Fourier modes). The universal enveloping algebra of the algebra of the oscillators is built but the Hopf algebra structure of the Heisenberg algebra is not correctly reproduced (the failure is at the level of the costructures). We showed that the problem is entirely solved and the complete Hopf algebra structure of the Heisenberg algebra is correctly induced if we adopt, simultaneously, Weyl ordering and Wigner's prescription of considering the Schr\"odinger oscillators as generators of a suitable Lie superalgebra, such as  $osp(1|2n)$. Taking into account the odd nature (opposed to the real, physical nature) of the oscillators, it is possible to obtain the Hopf algebra $\mathcal{U}(\mathfrak{h})$ and twist it into $\mathcal{U}^{\mathcal F}(\mathfrak{h})$. The deformation of $\mathcal{U}(\mathfrak{h})$ is totally compatible with the deformation of the algebra of the basic oscillators.

We finally investigated the twist deformations of the fermionic Heisenberg algebra and of the one-dimensional ${\cal N}$-extended supersymmetric quantum mechanics. We presented two possible constructions. First, for even values of ${\cal N}$, one can identify the supersymmetry algebra with the fermionic Heisenberg algebra and obtain the deformation by a simple change of symbols. Alternatively, one can adopt the superspace representation, in which the algebra of supersymmetric quantum mechanics is realized in terms of operators belonging to the universal enveloping algebra of one bosonic and several fermionic oscillators, and deform the supersymmetry algebra in this setting. We recovered, in a more general mathematical context, some cliffordization results already known the literature, both in terms of deformed generators and in terms of a deformed product defined on a module. We showed that, even if the twist is fermionic, the bosonic sector of the theory undergoes deformation in its multiparticle states, as can be inferred from the nontrivial deformation of the coproduct.

\end{document}